\documentclass[10pt, conference, compsocconf]{IEEEtran}

\usepackage{graphicx,amsmath, amsfonts}
\usepackage[english]{babel}
\usepackage[utf8]{inputenc}
\usepackage[T1]{fontenc}
 \usepackage{pifont}
\usepackage{rotating}
\usepackage{color}
 \usepackage{bbm}
 \usepackage{wasysym}
   \usepackage{stmaryrd}
\usepackage{cite}
\newlength\shadedboxwidth
\long\def\shaded#1{\begin{trivlist}\item[]%
\setlength\fboxsep{2ex}% 
\setlength\fboxrule{.4pt}% 
\def\bgcolor{.98}% 
\setlength\shadedboxwidth\linewidth
\addtolength\shadedboxwidth{-2\fboxsep}%
\addtolength\shadedboxwidth{-2\fboxrule}%
\fcolorbox[gray]{0}{\bgcolor}%
{\parbox{\shadedboxwidth}{#1}}\end{trivlist}}
 
\newtheorem{numer}{\hspace{-4mm}($\spadesuit$\hspace{-1mm}}

\newtheorem{tnumer}{\hspace{-4mm}($\clubsuit$\hspace{-1mm}}

\newcommand{\eop}{\hfill {$\Box$}}

\newtheorem{theorem}{\hspace{-3.5mm}Theorem}
 
 \newtheorem{lemmaa}[theorem]{\hspace{-3.5mm}Lemma}
 
 \newtheorem{definitionn}[theorem]{\hspace{-3.5mm}Definition}
\newtheorem{observ}[theorem]{\hspace{-3.5mm}Observation}

\newcommand{\lemmaaa}[1]{\vspace{1.3mm}{\em \begin{lemmaa} #1 \end{lemmaa} }\vspace{1.5mm}}
\newcommand{\definitionnn}[1]{\vspace{1.3mm} {\em \begin{definitionn} #1 \end{definitionn} } \vspace{1.5mm}}
\newcommand{\outline}[1]{
\noindent
\underline{~~~~~~~~~~~~~~~~~~~~~~~~~~~~~~~~~~~~~~~~~~~~~~~~~~~~~~~~~~~~~~~~~~~~~~~}\vspace{-1mm}\\
\noindent
\underline{~~~~~~~~~~~~~~~~~~~~~~~~~~~~~~{\sc outline}~~~~~~~~~~~~~~~~~~~~~~~~~~~~~~}\\

\vspace{-2mm} #1 \vspace{-2mm}

\noindent
\underline{~~~~~~~~~~~~~~~~~~~~~~~~~~~~~~~~~~~~~~~~~~~~~~~~~~~~~~~~~~~~~~~~~~~~~~~}\vspace{-5mm}\\
\noindent
\underline{~~~~~~~~~~~~~~~~~~~~~~~~~~~~~~~~~~~~~~~~~~~~~~~~~~~~~~~~~~~~~~~~~~~~~~~}\\
}

\newcommand{\spr}[2]{\text{\footnotesize\ding{72}}^{#1}_{\hspace{-0.5mm}#2}}
\newcommand{\spg}[2]{\text{\footnotesize\ding{73}}^{#1}_{\hspace{-0.5mm}#2}}

\newcommand{\fff}{\mathbbm{f}}

\newcommand{\oaa}{\curlywedgeuparrow }
\newcommand{\obb}{\curlyveeuparrow }

\author{Tomasz Gogacz, Jerzy Marcinkowski,  \\
Institute of Computer Science, University Of Wroclaw,\\
\\
January 8th, 2015
}

\title{The Hunt for a Red Spider:\\ Conjunctive Query Determinacy Is Undecidable.}

\begin{document}
\maketitle

\begin{abstract}
We solve a well known, long-standing open problem in relational databases theory, showing that the conjunctive query determinacy problem 
(in its ''unrestricted'' version) is undecidable.
 \end{abstract}
% 

%%%%%%%%%%%% Introduction %%%%%%%%%%%%%%%%%%%%%%%

\section{Introduction\\}

%%%%%%%%%%%%%%%%%%From Joinless to Sticky %%%%%%%%%%%%%%%%%

Imagine there is a database we have no direct access to, but there are views of this database available to us, defined by some 
conjunctive queries $Q_1$, $Q_2, \ldots Q_k$.  And we are given another conjunctive query $Q_0$. Will we be able to compute $Q_0$
only using the available views? The answer depends on whether the queries $Q_1$, $Q_2, \ldots Q_k$ {\em determine} query $Q_0$. 
To state it more precisely, the Conjunctive Query Determinacy Problem (CQDP)  is:

\shaded{
  The instance of the problem is a set of conjunctive queries ${\cal Q}=\{Q_1,\ldots Q_k\}$, and 
another conjunctive query $Q_0$.

 The question is whether ${\cal Q}$ determines $Q_0$, which means that for each two structures (database instances) ${\mathbb D}_1$ and ${\mathbb D}_2$ such that  
$Q({\mathbb D}_1)= Q({\mathbb D}_2)$ for each $Q\in \cal Q$, it also holds that  $Q_0({\mathbb D}_1) = Q_0({\mathbb D}_2)$.}

The  technical result of this paper is:\medskip

\begin{theorem}\label{main00}
CQDP is undecidable.\medskip
\end{theorem}

It is hard to imagine a more natural problem than CQDP, and better motivated. 
Answering queries using views appears in various contexts, 
see for example 
\cite{H01}  for a survey. Or \cite{DPT99}, where  the context is  query evaluation plans optimization. 
Or -- to see more recent examples -- \cite{FG12} where the view update problem is studied and 
 \cite{FKN13} where the context are description logics.  
It is fair to say that many variants of the problem are being considered, and the case we study, where both the views and the query are conjunctive queries,
is not the only possible scenario. But it is of special importance as the CQs -- as \cite{NSV07} puts it -- are 
''the simplest and most common language to define views and queries''

As we said it is hard to imagine a more natural problem than CQDP. 
So no wonder it has a 30 years long history as a research subject. And this history happens to be  quite complicated, marked by 
errors and confusion. 

The oldest paper we were able to trace, where CQDP is studied, is \cite{LY85}, whose first sentence is almost the same as ours:
{\em ''Assume that a set of derived relations is available in a stored form. Given a query, can it be computed from the derived relations and,
if so, how?''}. It was shown there, and in the next paper \cite{YL87}, by the same authors, 
that the problem is decidable if $\cal Q$ consists of just one query without self-joins
(there is however some additional form of selection allowed there, so it is not really comparable to the CQ paradigm).
Over the next 30 years 
many other decidable cases have been found. Let us just cite the most recent results: \cite{NSV10} shows that the problem is decidable if each  query from $\cal Q$ has only one free variable; in 
\cite{A11} decidability is shown for $\cal Q$ and $Q_0$ being ''path queries''. This is generalized in \cite{P11} to the the scenario where $\cal Q$ are path queries but $Q_0$ is any conjunctive
query. 

As we said in the Abstract, decidability of CQDP was a long standing open problem. It was indeed open, since 1985, but not without pauses. It was shown in 
\cite{LMS95} that it is decidable whether -- for given $\cal Q$ and $Q$ like in CQDP -- there exists another query $Q'$, over the signature consisting of 
$Q_1,Q_2,\ldots Q_k$, such that for each structure (database instance) $\mathbb D$ there is $Q_0({\mathbb D})=Q'(Q_1({\mathbb D}),\ldots Q_k({\mathbb D}))$ (notice that while 
the ''input'' of $Q$ are the relations of $\mathbb D$, which we do not have access to, the ''input'' of $Q'$ are the views that we are allowed to see). Existence of 
such $Q'$ 
-- a rewriting of $Q$ -- indeed implies determinacy. But -- and this fact was for a long time surprisingly poorly understood  -- not necessarily determinacy 
implies existence of a rewriting. There is no sign in \cite{LMS95} that the authors were aware of this distinction, and it seems that the first to realize 
that there is any problem here were the authors of \cite{SV05}. After realizing that conjunctive query determinacy and 
conjunctive query rewriting (as above defined) are possibly two different notions they show that they are in fact equivalent. Together with the
result of \cite{LMS95} this would imply decidability of CQDP.   But -- again surprisingly -- 
this proof was not correct, as spotted by (a superset of) the  authors of \cite{SV05} in \cite{NSV07}. Also in \cite{NSV07} a (correct) counterexample is presented,
of $\cal Q$ and $Q_0$ such that $\cal Q$ determines $Q_0$ but no rewriting $Q'$ being a CQ exists. In fact -- as it is also shown in  \cite{NSV07} -- 
$Q_0({\mathbb D})$ is not always a monotonic function of $Q_1({\mathbb D}), \ldots Q_k({\mathbb D})$.

Coming back to decidability of the determinacy problem: the paper  \cite{NSV07} is also the first to 
present a negative result. It was shown there, that the problem is undecidable if unions of conjunctive queries are allowed
rather than CQs. In \cite{NSV10} it was also proved that determinacy is undecidable if the elements of $\cal Q$ are CQs and $Q_0$ is a first order sentence 
(or the other way round). Another  negative result is presented in \cite{FGZ12}: determinacy is shown there to be undecidable  if  $\cal Q$  is a DATALOG program and 
$Q_0$ is CQ.

In our setting the instance of the problem consists of the set $\cal Q$ of the queries that define 
the views and of the query $Q_0$. A natural question to ask would be 
what happens if $Q_1({\mathbb D}),\ldots Q_k({\mathbb D})$ were also part of the input. This problem can be easily  shown to be decidable.
Complexity is studied  in \cite{AD98}.

\subsection{Finite vs. unrestricted case.} As usually in database theory there are two variants of the problem that one can consider: {\em finite}, where all the
structures in question (which in our case means ${\mathbb D}_1$ and ${\mathbb D}_2$) are assumed to be finite, and {\em unrestricted}, where there is no
such assumption. Most of the results of \cite{LMS95}, \cite{NSV07},  \cite{NSV10}, \cite{A11}  and \cite{P11} that we report above hold true regardless of the finiteness assumption. Unlike them, 
Theorem \ref{main00} of this paper concerns the unrestricted case only. Decidability of CQDP for the finite case remains open.

%%%%%%%%%%%%%%%%%%%%%%%%%%%%%%%%%%%%%%%%%%%%%%%%%%%%%%%%%%%%%%%%%%%%%%%%%%%%%%%%%%%%%%%%%%%%%%%%%%%%%%%%%%%%%%%%%%%%%%%%%%%%%%%%%%%
%%%%%%%%%%%%%%%%%%%%%%%%%%%%%%%%%%%%%%%%%%%%%%%%%%%%%%%%%%%%%%%%%%%%%%%%%%%%%%%%%%%%%%%%%%%%%%%%%%%%%%%%%%%%%%%%%%%%%%%%%%%%%%%%%%%
%%%%%%%%%%%%%%%%%%%%%%%%%%%%%%%%%%%%%%%%%%%%%%%%%%%%%%%%%%%%%%%%%%%%%%%%%%%%%%%%%%%%%%%%%%%%%%%%%%%%%%%%%%%%%%%%%%%%%%%%%%%%%%%%%%%

\outline{The rest of the paper is devoted to the proof of Theorem \ref{main00}.}

%%%%%%%%%%%%%%%%%%%%%%%%%%%%%%%%%%%%%%%%%%%%%%%%%%%%%%%%%%%%%%%%%%%%%%%%%%%%%%%%%%%%%%%%%%%%%%%%%%%%%%%%%%%%%%%%%%%%%%%%%%%%%%%%%%%
\section{Preliminaries}

In Section \ref{basic}  we  recall some standard finite model theory/database theory notions. They way we present them is rather standard. 
In  Sections \ref{tgds} and \ref{chase}   we also recall standard notions, but our notations may be seen as slightly non-standard (although of course equivalent to standard). 
This is how we think we need them in 
this paper. 

\subsection{Basic notions}\label{basic}

When we say ''structure'' we mean a relational structure $\mathbb  D$ over some signature $\Sigma$, i.e. a set of elements (vertices), denoted as $Dom({\mathbb D})$ and a set of relational atoms, whose arguments are elements of $\mathbb D$ and whose predicate names are from $\Sigma$.  Atoms are (of course) only positive. 
For an atomic formula $A$ we use notation ${\mathbb D}\models A$ to say that $A$ is an atom of $\mathbb D$. 

Apart from predicate symbols  $\Sigma$ can also contain constants. If $c$ is a constant from $\Sigma$ and $\mathbb D$ is a structure over $\Sigma$ then $c\in Dom({\mathbb D})$.

${\mathbb D}_1$ a substructure
of $\mathbb  D$ (and $\mathbb  D$ is a superstructure of  ${\mathbb D}_1$) if for each atom $A$ if  ${\mathbb  D_1}\models A$ then ${\mathbb  D}\models A$. This implies that  $Dom({\mathbb  D_1})\subseteq Dom({\mathbb  D})$.

For two structures ${\mathbb  D}_1$ and  $\mathbb  D$ over the same signature $\Sigma$ a function $h: Dom({\mathbb  D_1})\rightarrow Dom({\mathbb  D})$   is called a homomorphism 
if for each $P\in \Sigma$ of arity $l$ and  each tuple $\bar a\in Dom({\mathbb D})^l$ if ${\mathbb D_1}\models P(\bar a)$ then ${\mathbb D}\models P(\overline{h(a)})$ 
(where $\overline{h(a)}$ is a tuple of 
images of elements of $\bar a$). Notice that ${\mathbb D}_1$ a substructure
of $\mathbb  D$ if and only if identity is a homomorphism from ${\mathbb D}_1$ to $\mathbb  D$.

A conjunctive query (over $\Sigma$), in short CQ, is a conjunction of atomic formulas (over $\Sigma$) whose arguments are either variables or the constants from $\Sigma$, preceded by 
existential quantifier binding some of the variables. It is very important in this paper to distinguish between a conjunctive query and its quantifier-free part. We usually write 
$\Psi$ or $\Phi$ for a conjunction of atoms without quantifiers and $Q$ (possibly with a subscript) for conjunctive queries, so that we have something like:\vspace{-3mm}

$$Q(\bar x) = \exists \bar y  \;\;\Psi(\bar y,\bar x)$$\vspace{-5mm}

where  $\Psi(\bar y,\bar x)$ is a formula being a conjunction of atomic formulas and  $\bar x$ is a tuple of variables which are free in $Q$.

For a conjunction of atoms $\Psi$ (or for a CQ $Q(\bar x) = \exists \bar y \;\Psi(\bar y,\bar x)$) the canonical structure of $\Psi$, denoted as $A[\Psi]$, is the structure 
whose elements are all the variables and constants appearing in $\Psi$ and whose atoms are atoms of $\Psi$. It is useful to notice that 
for a structure $\mathbb D$ and a set $V\subseteq Dom({\mathbb D})$ there is a unique 
conjunctive query $Q$ such that ${\mathbb D}=A[Q]$ and that $V$
is the set of free variables of $Q$. 

For a CQ $Q(\bar x) = \exists \bar y \;\Psi(\bar y,\bar x)$ with $\bar x=x_1, \ldots x_l$, a  structure $\mathbb D$ and a tuple $a_1,\ldots a_l$ of elements of $\mathbb D$ 
we write ${\mathbb D}\models Q(a_1,\ldots a_l)$ when there exists a homomorphism $h: A[\Psi] \rightarrow \mathbb D$ such that $h(x_i)=a_i$ for each $i$. 

Sometimes we also write  ${\mathbb D}\models Q$. Then we assume that all the free variables of $Q$ are implicitly existentially quantified, so that the meaning of the notation is 
that there exists any  homomorphism $h: A[\Psi] \rightarrow \mathbb D$.

The most fundamental  definition of this paper now, needed to formulate the problem we solve:
 for a CQ $Q$ and for a structure $\mathbb D$ by $Q({\mathbb D})$ we denote the ''view defined by $Q$ over $\mathbb D$'', which is the relation
$\{\bar a: {\mathbb D}\models Q(\bar a)\}$. 

\subsection{TGDs and how they act on a structure}\label{tgds}

A Tuple Generating Dependency (or TGD) is a formula of the form:\vspace{-5mm}

$$\forall \bar x, \bar y\; \Phi(\bar x, \bar y)\Rightarrow \exists \bar z \; \Psi(\bar z,\bar y)$$\vspace{-5mm}

where $\Psi$ and $\Phi$ are -- as always -- conjunctions of atomic formulas. The standard convention, which we will obey, is that 
the universal quantifiers in front of the TGD are omitted.

From the point of view of this paper it is important to see a TGD -- let it be $T$, equal to $\Phi(\bar x, \bar y)\Rightarrow \exists \bar z \; \Psi(\bar z,\bar y)$ --
 as a procedure whose input is a structure $\cal D$ and whose output is a new structure 
being a superstructure of  $\cal D$:\smallskip

{\tt

find a tuple $\bar b$ (with 
% $|\bar a|= |\bar x|$ and
 $|\bar b|= |\bar y|$) such that:

\ding{182} ${\cal D}\models \exists \bar x\; \Phi(\bar x, \bar b)$ via homomorphism $h$ but

\ding{183} ${\cal D}\not\models \exists z\; \Psi(\bar z, \bar b)$;\smallskip

create a new copy  of $A[\Psi]$;\smallskip

output  $T({\cal D}, \bar b)$ being a union of $\cal D$ and the new copy of $A[\Psi]$, with each $y$ from $A[\Psi]$  identified with $h(y)$ in $\cal D$.\medskip 
}

The message,  which will  be {\bf good to remember}, is that the interface between the ''new'' part of the structure, added by a single application of a TGD to 
a structure, and the ''old'' structure, are the free variables of the query in the right  hand side of the TGD. 

\subsection{Chase and its universality}\label{chase}

For a structure $\cal D$ and a set $\cal T$ of TGDs let ToDo$({\cal T},{\cal D})$ be the set of all  pairs $\langle \bar b,  T \rangle $ such that 
 $T$, equal to $\Phi(\bar x, \bar y)\Rightarrow \exists \bar z \; \Psi(\bar z,\bar y)$, is a  TGD from $\cal T$ and $\bar b$ satisfies conditions \ding{182} and \ding{183}. Roughly speaking ToDo$({\cal T},{\cal D})$ is the set of tuples of elements of $\cal D$ which satisfy the 
left hand side of some TGD in $\cal T$
but still wait for a witness -- confirming that they also satisfy the right hand side -- to be added to the structure. 

A sequence $ \{{\cal D}_i\}_{i\in \Omega} $ of structures, for some ordinal number $\Omega$, 
will be called {\em fair} (with respect to $\cal T$ and ${\mathbb D}$) if:

\noindent\textbullet~~ ${\cal D}_0={\mathbb D}$;

\noindent\textbullet~~
for each $i>0$ we have $ {\cal D}_{i} = T(\bigcup_{j<i}{\cal D}_j, \bar b) $ for some   
$\langle  \bar b, T\rangle \in$ ToDo$({\cal T},\bigcup_{j<i}{\cal D}_j)$;

\noindent\textbullet~~
for each $\langle \bar b, T \rangle \in$ToDo$({\cal T},\bigcup_{j<i}{\cal D}_j)$ for some $i$, there is $k>i$ such that 
$\langle \bar b, T \rangle \not\in$ToDo$({\cal T},\bigcup_{j<k}{\cal D}_j)$.\smallskip

Let $ \{{\cal D}_i\}_{i\in \Omega} $ be a fair sequence (with  respect to $\cal T$ and ${\mathbb D}$).
 Then the structure  $Chase({\cal T},{\mathbb D})$ is defined as $\bigcup_{i\in \Omega}{\cal D}_i$ and each of the sets ${\cal D}_i $ is
called a {\em stage of} $Chase$. 

In other words, $Chase({\cal T},{\mathbb D})$ is a structure being result of adding, one by one, tuples that witness that some TGD from $\cal T$ is satisfied  
for a given tuple from the current structure. The set ToDo always contains tuples that do not have the required witnesses yet. Notice that there are two 
possible ways, for  a tuple 
$\langle \bar b, T \rangle$, to disappear from the set ToDo: one is that the TGD $T$ is applied to the tuple $\bar b$ at some step. But it may also happen that 
the witnesses $\bar b$ needs are added as a side-effect of other TGDs being applied to other tuples\footnote{A reader who is aware of the difference between standard and oblivious Chase will notice that what we define is the standard/lazy version.}.

It may appear, and not without a reason, that the structure $Chase({\cal T},{\mathbb D})$ depends on the ordering in which tuples are selected from the ToDo set. 
But the beautiful fact (and a well-known one, \cite{JK82}) is that, regardless of the ordering:

~

\begin{theorem}[Chase as universal structure]\label{universal}

\noindent\textbullet~~ $Chase({\cal T},{\mathbb D})\models \cal T$. In other words if $\Phi(\bar x, \bar y)\Rightarrow \exists \bar z \; \Psi(\bar z,\bar y)$ is a TGD from $\cal T$ 
and  $Chase({\cal T},{\mathbb D})\models \exists \bar x\; \Phi(\bar x, \bar b)  $ then also  $Chase({\cal T},{\mathbb D})\models  \exists z\; \Psi(\bar z, \bar b)$ 

\noindent\textbullet~~ Let $\mathbb M$ be any superstructure of ${\mathbb D}$ such that ${\mathbb M}\models \cal T$ and let $Q$ be any conjunctive query such that 
 $Chase({\cal T},{\mathbb D})\models Q$. 
Then also   ${\mathbb M}\models Q$. 
\end{theorem}

~

Most of the lemmas in this paper,
concerning the structure of $Chase({\cal T},{\mathbb D})$  for specific $\cal T$ and $\mathbb D$ 
  are proved by induction on a respective fair sequence, even if this is 
not always mentioned explicitly.

\subsection{Thue systems}\label{thuesystems}

Our undecidability proof is by reduction from a variant of the Thue systems word problem (also known as semigroups word problem). A Thue system is 
given by a finite symmetric relation $\Pi\subseteq {\cal A}^* \times {\cal A}^*$ for some finite alphabet $\cal A$. For two words ${\bf w, w'}\in {\cal A}^*$ we define
 ${\bf w} \stackrel{}{\Leftrightarrow_\Pi} {\bf w'}$ if and only if there are words ${\bf v, v'}\in {\cal A}^*$ and a pair $\{{\bf t, t'}\}\in \Pi$ such that
${\bf w}={\bf vtv'}$ and ${\bf w'}={\bf vt'v'}$. Relation $\stackrel{\ast\;\;\;}{\Leftrightarrow_\Pi}$ is defined as the transitive closure of $\stackrel{}{\Leftrightarrow_\Pi}$.
Various undecidability results involving  relation  $\stackrel{\ast\;\;\;}{\Leftrightarrow_\Pi}$ can be proved using standard techniques from \cite{D77}.

%%%%%%%%%%%%%%%%%%%%%%%%%%%%%%%%%%%%%%%%%%%%%%%%%%%%%%%%%%%%%%%%%%%%%%%%%%%%%%%%%%%%%%%%%%%%%%%%%%%%%%%%%%%%%%%%%%%%%%%%%%%%%%%%%%%

\section{Green-Red TGDs}\label{g-r}

\subsection{Green-Red Signature}

For a given signature $\Sigma$ let $\Sigma_G$ and $\Sigma_R$ be two  copies of $\Sigma$ having new relation symbols, which have the same names and the same arities as 
symbols in $\Sigma$ but are written in green and red respectively. Let $\bar\Sigma$ be the union of   $\Sigma_G$ and $\Sigma_R$.
Notice that the constants from $\Sigma$, not being relation symbols, are never colored and thus survive in $\bar\Sigma$ unharmed. 
 
For any formula $\Psi$ over $\Sigma$ let $R(\Psi)$ (or $G(\Psi)$) be the result of painting all the predicates in $\Psi$ red (green). 
For any formula $\Psi$ over $\bar\Sigma$ let $dalt(\Psi)$ (''daltonisation of $\Psi$'') be a formula over $\Sigma$ being  the result of erasing the colors from 
predicates of $\Psi$. The same convention applies to structures. Whenever an uncolored relation symbol (usually {\small H}) is used 
in the context of  $\bar \Sigma$  it should be understood as ''G({\small H}) or R({\small H})''.

\subsection{Having ${\mathbb D}$ instead of ${\mathbb D}_1$ and  ${\mathbb D}_2$.}

We prefer to restate CQDP a little bit in order to be talking about one database instance instead of two. Clearly CQDP is equivalent to:\smallskip

\shaded{
\noindent
{\bf The green-red conjunctive query determinacy problem (GRCQDP).} The instance of the problem is a finite 
set $\cal Q$  of conjunctive queries 
% $Q_1, Q_2 \ldots Q_l$,
 and 
another conjunctive query $Q_0$, all of them over some signature $\Sigma$. 
The question is whether for each structure ${\mathbb D}$ over  $\bar\Sigma$ such that:\smallskip

\ding{184} $(G(Q))({\mathbb D})= (R(Q))({\mathbb D})$ for each $Q\in \cal Q$\smallskip

it also holds that $(G(Q_0))({\mathbb D})= (R(Q_0))({\mathbb D})$.}\medskip

For a conjunctive query $Q$ of the form $\exists \bar x \; \Phi(\bar x,\bar y)$ where $\Phi$ is a conjunction of atoms over $\Sigma$ let 
$Q^{G\rightarrow R}$  be the TGD {\em generated by} $Q$ in the following sense:\vspace{-4mm}

 $$Q^{G\rightarrow R}\;\;\;=\;\;\;\forall \bar x, \bar y \;[\; G(\Phi)(\bar x,\bar y) \Rightarrow \exists \bar z \; R(\Phi)(\bar z,\bar y)\;]$$\vspace{-4mm}

TGD $Q^{R\rightarrow G}$ is defined in an analogous way. For a set $\cal Q$ as above let ${\cal T}_{\cal Q}$ be the 
set of all TGDs of the form $Q^{G\rightarrow R}$ or $Q^{R\rightarrow G}$ with $Q\in \cal Q$.
It is very easy to see that:

\lemmaaa{
The above condition \ding{184} is satisfied by structure $\mathbb D$ if and only if ${\mathbb D}\models {\cal T}_{\cal Q}$.
}

Now  GRCQDP can be again restated as:

\shaded{
Given a set $\cal Q$ (as in the original formulation of GRCQDP, above), and 
another conjunctive query $Q_0$, 
 is it true that:\smallskip

\ding{185} for  each structure $\mathbb D$ and each tuple $\bar a$ of elements of $\mathbb D$,
if ${\mathbb D}\models {\cal T}_{\cal Q}, G(Q_0)(\bar a)$ then  also  ${\mathbb D}\models R(Q_0)(\bar a)$\hfill ?}

But \ding{185}  means that ${\cal T}_{\cal Q}, G(Q_0)(\bar a)\models  R(Q_0)(\bar a)$
where $\bar a$ is a tuple of new constants. Thus -- by Theorem \ref{universal} -- CQDP is equivalent to\footnote{
The observation that determinacy can be semi-decided using chase is not ours and is at least as old as \cite{NSV07}. The difference is that in \cite{NSV07} they 
prefer to see two separate structures rather than two colors.} 
\shaded{
{\bf CQDP -- the green-red Chase version (CQDP-GRC).} Given the set $\cal Q$ 
(as in the original formulation, above), and 
another conjunctive query $Q_0$, 
 is it true that:\smallskip

$ Chase({\cal T}_{\cal Q}, A[G(Q_0)(\bar a)])\models  R(Q_0)(\bar a)$\hfill ?}

where $A[G(Q_0)(\bar a)]$  is the canonical structure of  $G(Q_0)(\bar a)$.\medskip

The {\bf main technical result of this paper} is:\medskip

\begin{theorem}[Theorem \ref{main00} restated]\label{main}
CQDP-GRC is undecidable.\medskip
\end{theorem}

 Of course the problem to determine, for given set $\cal T$ of TGDs,  database instance $\mathbb D$ and query $Q$, whether $Chase({\cal T},{\mathbb D})\models Q$,  is undecidable in general. 
But this does not {\em a priori} mean that CQDP-GRC is undecidable, since the TGDs we allow here are of very special green-red form (with the head being just recoloring of the body)  and
since we only consider $Q$ being a recoloring of $\mathbb D$.

\subsection{Idempotence lemma}

One useful feature of the green-red TGDs is described in the following easy lemma:

~

\lemmaaa{\label{idempotencja}
Let $\cal Q$ be a set of conjunctive queries and let the set ${\cal T}_{\cal Q}$ of the green-red TGDs generated by $\cal T$  be defined as before. 
Let $T$ be  $Q^{R\rightarrow G}$ for some $Q\in \cal Q$ and
suppose $\bar b\in \cal D$ is such that $\langle \bar b, T  \rangle\in\;$ToDo$({\cal T}_{\cal Q},{\cal D})$. Suppose 
  ${\cal D}'= T({\cal T}, \bar b)$.

Then  $\langle \bar b, Q^{G\rightarrow R}  \rangle\not\in\;$ToDo$({\cal T}_{\cal Q},{\cal D}')$
}

\noindent {\em Proof:} The necessary condition for $ \langle \bar b, Q^{R\rightarrow G}  \rangle $ to be in ToDo$({\cal T}_{\cal Q},{\cal D})$ is that 
${\cal D}\models R(Q)(\bar b)$. Since ${\cal D}'$  is a superstructure of ${\cal D}$ we also have ${\cal D'}\models R(Q)(\bar b)$. 
But the necessary condition for $ \langle \bar b, Q^{G\rightarrow R}  \rangle $ to be in ToDo$({\cal T}_{\cal Q},{\cal D})$ is that 
${\cal D'}\not\models R(Q)(\bar b)$.\eop

Of course both the lemma and its proof also hold for the colors reversed.

\outline{\noindent The rest of this paper is devoted to the proof of of Theorem \ref{main}.
The proof  is by encoding the word problem for some very specific Thue systems over a very specific alphabet (being a subset of) ${\mathbb A}_s$.

In Section \ref{monadyczne} we study s-piders, which are elements of the set ${\mathbb A}_s$, and 
s-pider queries  ${\mathbb F}_s$ which are partial functions from ${\mathbb A}_s$ to ${\mathbb A}_s$. 
 
Then, in Section \ref{binarne} and later, we show how to concatenate s-piders into words, and how 
to modify ${\mathbb F}_s$ to get functions that take a pair of elements of ${\mathbb A}_s$ as an input, and 
output pairs of elements of ${\mathbb A}_s$. This opens the way to  Thue systems encoding.}

%%%%%%%%%%%%%%%%%%%%%%%%%%%%%%%%%%%%%%%%%%%%%%%%%%%%%%%%%%%%%%%%%%%%%%%%%%%%%%%%%%%%%%%%%%%%%%%%%%%%%%%%%%%%%%%%%%%%%%%%%%%%%%%%%%%
%%%%%%%%%%%%%%%%%%%%%%%%%%%%%%%% %%%%%%%%%%%%%%%%%%%%%%%%%%

\section{S-piders and graph reachability}\label{monadyczne}

Let $s\in \mathbb N$ be fixed and let $\Sigma$  be a signature consisting of:

 \begin{enumerate}
\item[--] constants $c_1, c_2, \ldots c_s$ and  $c^1, c^2, \ldots c^s$

\item[--] binary relation symbols $C_1, C_2, \ldots C_s$ and  $C^1, C^2, \ldots C^s$ (the $C$ reads as ''calf'' here)

\item[--] binary relation symbols $T_1, T_2, \ldots T_s$ and  $T^1, T^2, \ldots T^s$  (the $T$ reads as ''thigh'')
       
\item[--] ternary relation symbol {\small H}.

\end{enumerate}

$\Sigma_G$, $\Sigma_R$ and  $\bar\Sigma$ are defined as in Section \ref{g-r}. 

For an element $a$ of a structure $\mathbb D$ 
over $\bar\Sigma$ by out-degree of $a$ we mean the number of atoms $P(a,b)$, with $P\in\bar\Sigma$ and $b\in\mathbb D$, which are 
true in $\mathbb D$. The in-degree  is defined in an analogous way. By  out-degree of $a$ 
with respect to $P$, with $P\in\Sigma\cup\bar\Sigma$ we mean the number of atoms $P(a,b)$, with $b\in\mathbb D$, which are 
true in $\mathbb D$.

From now on $i,j$ are {\bf always} natural numbers from the set ${\mathbb S}= \{1,2,\ldots s\}$. 
Another notation we use is $I,J\subseteq {\mathbb S}$ which {\bf always} mean either a singleton or the empty set. Being computer scientists, we do not distinguish between
a singleton and its only element.

\subsection{S-piders and their taxonomy}\label{taxonomy}

For a conjunction of atomic formulas $\Psi$ and for an atom $P$ (atoms $P,P'$) occurring in $\Psi$ let 
$\Psi / P$ (resp. $\Psi / P, P'$) be $\Psi$ with $P$ (resp. $P$ and $P'$) removed from the conjunction.

Let now $\Phi_s$ be defined as the following conjunction of atomic formulas:\smallskip

\noindent
{\small H}$(z,z_1,z_2)\wedge \bigwedge_{i=1}^{s} T_i(z,x_i)\wedge T^i(z,y_i) \wedge C_i(x_i,c_i)\wedge C_i(y_i,c^i) $\smallskip

\definitionnn{
\noindent
\textbullet~
The {\em  ideal green full s-pider}, denoted as $\spg{}{}$, is  $A[G(\Phi_s)]$ --  the canonical structure of 
the green version of $\Phi_s$. The {\em  ideal red full s-pider} $r$, denoted as $\spr{}{}$,  is   $A[R(\Phi_s)]$.\medskip

\noindent
\textbullet~
An {\em ideal green 1-lame  upper  s-pider}, denoted $\spg{i}{}$, is 
$A[G(\Phi_s/C^i(y_i,c^i))\wedge R(C^i(y_i,c^i))]$. An {\em ideal red 1-lame  upper  s-pider},
denoted $\spr{i}{}$, is 
$A[R(\Phi_s/C^i(y_i,c^i))\wedge G(C^i(y_i,c^i))]$.\medskip

\noindent
\textbullet~
An {\em ideal green 1-lame  lower  s-pider}, denoted $\spg{}{i}$, is 
$A[G(\Phi_s/C_i(x_i,c_i))\wedge R(C_i(x_i,c_i))]$. An  {\em ideal red 1-lame  lower  s-pider},
denoted $\spr{}{i}$, is 
$A[R(\Phi_s/C_i(x_i,c_i))\wedge G(C_i(x_i,c_i))]$.\medskip

\noindent
\textbullet~
An {\em ideal green 2-lame  s-pider},   denoted $\spg{i}{j}$, is
$A[G(\Phi_s/C_i(x_i,c_i), C^j(y_j,c^j))\wedge R(C_i(x_i,c_i)\wedge C^j(y_j,c^j))]$.\medskip

\noindent
\textbullet~
An {\em ideal red 2-lame  s-pider},   denoted $\spr{i}{j}$, is 
$A[R(\Phi_s/C_i(x_i,c_i), C^j(y_j,c^j))\wedge G(C_i(x_i,c_i)\wedge C^j(y_j,c^j))]$.
}

Notice that each of ideal s-piders really looks exactly like a spider: there is a {\em head} ($z$), with $2s$ legs attached to it; each leg has length 2, 
and the legs are distinguishable. Head is connected to the {\em tail} ($z_1$) and the {\em antenna}
 ($z_2$). But the antenna and the tail  will not bother us in this Section.

Full s-piders -- red and green -- are monochromatic, head and all legs must be of the same color. 1-lame s-piders have one calf of the opposite color. As we distinguish between the ''upper'' and ''lower'' legs of a s-pider, we have two kinds of 1-lame s-piders of each color. A 2-lame s-pider has one upper calf and one lower calf of the 
opposite color. Any  of the $2s$ vertices of a s-pider which are  neither head nor a constant will be called {\em a knee}. Sometimes we will need to be more precise, and talking about particular s-pider 
we will use descriptions like
''$i$'th upper knee'',  hoping that meaning of it is clear.

\definitionnn{
${\mathbb A}_s$ is the set of all $\spr{I}{J}$ and $\spg{I}{J}$, with $I$ and $J$ as defined above.
A s-pider $\spr{I}{J}$ (or $\spg{I}{J}$) is called {\em upper} if $I$ is non-empty and is called   {\em lower} if $J$ is non-empty. 
}

In other words ${\mathbb A}_s$ is the set of all ideal s-piders: full, 1-lame and 2-lame, both red and green.
Notice that a 1-lame s-pider is always either upper or lower, a 2-lame s-pider is both, and a full s-pider is neither upper nor lower.

While ideal s-piders are finitely many ($2+4s+2s^2$ of them), for each structure over $\bar\Sigma$ there can be 
many --  maybe infinitely many -- actual incarnations of ideal s-piders in this structure:

\definitionnn{
A {\em real s-pider} is any structure $\cal S$ (in particular a substructure of another structure) such that:
\begin{itemize}
\item 
$dalt({\cal S})\models \Phi_s$,

\item 
if ${\cal S}'$ is a proper substructure of ${\cal S}'$ then $dalt({\cal S}')\not\models \Phi_s$.
\end{itemize}
}

The second condition of the above definition looks more complicated than it really is. We just do not want 
a house full of s-piders to be called a s-pider.

\subsection{S-pider queries and what they are good for.}

Let us  first remind the reader that for each structure $\mathbb D$, and each subset  $V$ of 
$Dom({\mathbb D})$ there exists 
a unique conjunctive query $\Psi$ such that $V$ is the set of free variables of $\Psi$ and   ${\mathbb D} = A[\Psi]$:

\definitionnn{[s-pider queries]

\begin{enumerate}

\item 
$\fff^i_j$ is the unique query with free variables $x_j$ and $y_i$ whose canonical structure is 
equal to $A[\Phi_s/C_j(x_j,c_j), C^i(y_i,c^i)]$;\smallskip

\item 
$\fff^i$ is the unique query with single free variable $y_i$ whose canonical structure is 
equal to 
$A[\Phi_s/C^i(y_i,c^i)]$\smallskip

\item
$\fff_i$ is the unique query with single free variable $x_i$ whose canonical structure is 
equal to 
$A[\Phi_s/C_i(x_i,c_i)]$\smallskip

\end{enumerate}
}

By analogy with s-piders, the s-pider queries of the form $\fff^i_j$ will be sometimes called 2-lame, 
and of the form $\fff^i$ or $\fff_j$ will be called 1-lame. 
And, like 1-lame s-piders, also 1-lame s-pider queries are either upper and lower, while 2-lame are both. 
 Let ${\mathbb F}_s$ be the set of all s-pider queries.
Let us now learn -- by examples -- how the green-red TGDs generated by the queries from ${\mathbb F}_s$ act on elements of ${\mathbb A}_s$.\medskip

\noindent
{\bf Example 1.} Suppose ${\cal Q}$ consist of a single query $\fff^i_j$ for some $i,j$ and let ${\cal T}_{\cal Q}$ be the set of TGDs, as defined in Section \ref{g-r}.
Let us try to understand how the TGDs of ${\cal T}_{\cal Q}$ can be applied to $\spr{i}{}$.

${\cal T}_{\cal Q}$ consists of two TGDs. One of them is  $(\fff^i_j)^{R\rightarrow G}$. 

It tries to find, in the current structure, a homomorphic image $\cal D$ of $A[R(\fff^i_j)])$ and, if this succeeds,
it: 
\begin{itemize}
\item[--] produces a fresh copy of $A[G(\fff^i_j)]$ and

\item[--] identifies elements of this copy resulting from free variables of $G(\fff^i_j)$ with 
  elements of $\cal D$ resulting from the respective free variables\footnote{Of course the constants from the language are seen as free variables here, 
and their different occurrences  are also identified.}  in $R(\fff^i_j)$.

\end{itemize}

   The other TGD,  $(\fff^i_j)^{G\rightarrow R}$,  does the same, but with the colors reversed.

Now, if the current structure is $\spr{i}{}$, which is red, then of course the only 
possible match is with $(\fff^i_j)^{R\rightarrow G}$. The s-pider $\spr{i}{}$ is lame, it lacks his
upper $i$-th 
calf\footnote{Actually it has one, but green, and the atoms in the body of any TGD of the form $Q^{R\rightarrow G}$ are red, so they only can match with red atoms.}
but it is not needed for a match since $G(\fff^i_j)$ lacks this calf too.

Thus a new -- green -- copy $G(A[(\fff^i_j)])$ of $A[(\fff^i_j)]$ will be created. How will it be connected
to the original $\spr{i}{}$~?

Of course all the constants  from $G(A[\fff^i_j])$ (which means all constants from $\Sigma$ apart from $c^i$ and $c_j$) 
will be identified with the respective constants in $\spr{i}{}$. Also the $i$-th upper knee of $G(A[\fff^i_j])$ will be identified with the respective knee of $\spr{i}{}$  and 
the $j$-th lower   knee of $G(A[\fff^i_j])$ will be identified with the respective knee of $\spr{i}{}$. Notice that,
while $G(A[\fff^i_j])$ is not a s-pider (it is two calves short of being one), we actually created a new s-pider. It consists of the copy
of $G(A[\fff^i_j])$, and of two calves that it shares with $\spr{i}{}$:  the $i$-th upper calf of $\spr{i}{}$ (which is green) and of the lower $j$-th calf of $\spr{i}{}$
 (which is red, and is the only red calf of the new s-pider). Not only we created a new s-pider but we already
have a name for it -- it is a copy of $\spg{}{j}$! We cannot resist the temptation of writing this as: \vspace{-4mm}

$$\fff^i_j(\spr{i}{})=\spg{}{j}$$\vspace{-4mm}

\noindent
{\bf Example 2.} Let now ${\cal Q}$ consist of a single query $\fff^i_j$ and consider a s-pider $\spg{}{k}$  with $k\neq j$.
Since $\spg{}{k}$ is green, there is of course no match with the left hand side of the TGD $(\fff^i_j)^{R\rightarrow G}$. But is there a match with the left hand side of
$(\fff^i_j)^{G\rightarrow R}$~? Notice that the atom $G(C_k(x_k,c_k))$ occurs in $G(\fff^i_j)$. But not in $\spg{}{k}$ -- the $k$'th lower calf of $\spg{}{k}$ is red.
We cannot resist the temptation of writing this as: \vspace{-2mm}
 
$$\text{if~~} k\neq j \text{~~then~~} \spg{}{k} \not\in Dom(\fff^i_j) $$\vspace{-4mm}

\noindent
{\bf Example 3.} Let again ${\cal Q}$ be the single query $\fff^i_j$ for some $i,j$.  We already know that $\fff^i_j(\spr{i}{})=\spg{}{j}$ or, to be more 
precise, that $Chase({\cal T}_{\cal Q}, \spr{i}{})\models \spg{}{j}$.
By exactly the same argument we get that:
\begin{itemize}
\item $Chase({\cal T}_{\cal Q}, \spr{}{j})\models \spg{i}{}$,
\item $Chase({\cal T}_{\cal Q},  \spg{i}{})\models  \spr{}{j}$,
\item $Chase({\cal T}_{\cal Q}, \spg{}{j})\models \spr{i}{}$.
\end{itemize}

\definitionnn{
Suppose ${\cal S},{\cal S}'\in {\mathbb A}_s$ are such that   ${\cal S}\neq{\cal S}'$. Let $X\in \{G,R\}$  be the of color of $\cal S$ and let $Y$ be the opposite color.
Let $T$ be the TGD: $ X(f)(\bar w,\bar u) \Rightarrow \exists \bar v \; Y(f)(\bar v, \bar u)$.

Then by  $f({\cal S})={\cal S}'$ we will mean that:

\noindent
\textbullet~
${\cal S}\models (X(f))(\bar b)$ for some tuple $\bar b$ of elements of ${\cal S}$;

\noindent
\textbullet~
${\cal S}'$ is a substructure of $T({\cal S},\bar b)$ 
}

In other words, $f({\cal S})={\cal S}'$ means that one of the two green-red TGDs generated by $f$ can be applied to $\cal S$ and that 
a copy of ${\cal S}'$ is then produced in one step. It is of course easy to see that the color of ${\cal S}'$ is then $Y$.

The  examples of the previous subsection can be easily extended to a proof of:

\lemmaaa{[Algebra of s-piders]\label{algebra}
Let $I,J,I',J'\subseteq \mathbb S$ be as before.
Then $\fff^I_J(\spr{I'}{J'})$ is defined  if and only if $I'\subseteq I$ and $J'\subseteq J$.
If this the case then $f^I_J(\spr{I'}{J'})=\spg{I\setminus I'}{J\setminus J'}$. 

The same holds for the colors reversed. 
}

\subsection{Example: Encoding graph reachability}\label{grosiagalnosc}

As one more  toy example, consider an undirected graph $\langle V,E\rangle $ , with $V=\{v_1,v_2,\ldots v_t\}$ and $E=\{e_1,e_2,\ldots e_{t'}\}$.
Suppose -- for simplicity of 
presentation -- that  degree of $v_1$ is  exactly 1, and that $e_1$ is the only edge containing $v_1$.
  
Let $s\in \mathbb N$ be such that $s\geq t$ and $s\geq t'$
and let the set $\cal Q$ contain the following s-pider queries: $\fff_1$, $\fff^2$ and, 
for each triple $i,j,k$ such that $e_k=\{v_i,v_j\}$, two queries: $\fff^i_k$ and $\fff^j_k$.

Now we can represent graph reachability as an instance of GRCQDP:

\begin{observ}\label{osiagalnosc}
The two conditions are equivalent:

\begin{itemize}
\item[(i)] There is a path, in $\langle V,E\rangle $, from $v_1$ to $v_2$;

\item[(ii)] $Chase({\cal T}_{\cal Q}, \spg{}{})\models \spr{}{}$

\end{itemize}
\end{observ}

For the proof\footnote{Remember, this is an example, so the goal is to see the mechanisms rather than a rigorous proof.}, suppose that there is a path $v_1,e_1,v_{i_1},e_{i_1}\ldots v_{i_l},e_{i_l},v_2$ from $v_1$ to $v_2$. One can see that 
$Chase({\cal T}_{\cal Q}, \spg{}{})\models \spr{}{}$ contains the following s-piders: $\spr{}{1}$ (produced from  $\spg{}{}$ by $\fff_1$),
$\spg{i_1}{}$ (produced from  $\spr{}{1}$  by $\fff_1^{i_1}$), $\spr{}{i_1}$, and so on. For each vertex $v_k$ reachable from $v_1$ the green 1-lame upper s-pider 
$\spg{k}{}$ will be at some point added to the chase and for each edge $e_k$ reachable from $v_1$ the red 1-lame lower s-pider 
$\spr{}{k}$ will be added. Finally, once we have $\spr{2}{}$, the query $\fff^2$ can be used to produce $\spr{}{}$.

But wait: how about the opposite direction?
 Clearly, the queries of $\cal Q$ were designed to 
only produce the red s-spiders for reachable edges and green for reachable vertices (as above) but how are we sure that there are no
side-effects leading to the creation  of $\spr{
}{}$ even if $v_2$ is not reachable from $v_1$? There could be two sources of such side-effects. 
One is that -- due to the complicated structure of $Chase({\cal T}_{\cal Q}, \spg{}{})$ new real s-piders could emerge there\footnote{
To be more precise, what we really fear here are not new s-piders but new -- unintended -- matchings with left hand side of some TGD from ${\cal T}_{\cal Q}$. See Lemma 
\ref{1spider} (iv).
}, 
which were not produced as $f({\cal S})$,
for $f\in \cal Q$ and $\cal S$ previously in $Chase({\cal T}_{\cal Q}, \spg{}{})$. This could in principle happen, the s-piders share constants, and knees and who knows what more.

Second possible source of problems is that some weird application of queries from $\cal Q$ to the s-piders we ourselves produced
could lead to creation of something more than just the representations of reachable vertices and edges (as described above).

As it turns out -- and as we are going to show before the end of this Section -- there are no side-effects of the first sort and while 
there indeed are some side-effects of the second sort,
but they are ''sterile'' and thus controllable.

\subsection{Understanding the structure of Chase(${\cal T}_{\cal Q}, \spg{}{}$)}

We want to make sure that our abstraction of low-level structures, like s-piders and TGDs, as high-level objects,
as symbols ${\mathbb A}_s$ and partial functions $f^I_J: {\mathbb A}_s \rightarrow {\mathbb A}_s$ is correct, in the sense 
that there are no uncontrollable side-effects. And this is what the following series of lemmas is about.

Let ${\cal Q}\subseteq {\mathbb F}_s$.  We are going 
to analyze the structure of $Chase({\cal T}_{\cal Q},\spg{}{} )$. Let 
Let $ \{ {\cal D}_k\}_{k\in\Omega}$ be a fair sequence (with respect to ${\cal T}_{\cal Q}$ and $\spg{}{}$)
 and recall (see Section \ref{chase}) that $Chase({\cal T}_{\cal Q}, \spg{}{})$ is defined as $\bigcup_{k\in \Omega} {\cal D}_k$.
Our basic proof technique will be induction on $k$.

\lemmaaa{\label{outdegree1}
Each knee in $Chase({\cal F}_{\cal Q}, \spg{}{})$ has out-degree 1. Each red head has out-degree 1 with respect to any red $T_i$ and to any red $T^i$ and
out-degree 0 with respect to any other relation. The same is true, with colors reversed, for green heads.
}

\noindent {\em Proof:} 
Induction. For the induction step notice that 
 atoms of relations $C_i$, $C^i$, $T_i$, $T^i$  can only be created by the TGDs from ${\cal T}_{\cal Q}$ together with 
their leftmost  argument. This means that an application of a rule  from ${\cal T}_{\cal T}$ can never add an out-going edge to an already existing 
element (notice that this is not true about in-coming edges, and this why s-piders can share a calf).\eop

\lemmaaa{[No low-level side-effects]\label{1spider}
Suppose an element $a$ in  $Chase({\cal T}_{\cal Q},\spg{}{})$ is  such that 
$Chase({\cal T}_{\cal Q}, \spg{}{})\models${\small H}$(a,a_1,a_2)$, for some $a_1$,$a_2$. Then:

\noindent
(i) There exists exactly one real s-pider $\cal S$ in $Chase({\cal T}_{\cal Q}, \spg{}{})$ such that $a$ is the head of $\cal S$.

\noindent
(ii) ${\cal S}$ is created together with $a$, which means that if ${\cal D}\in \{ {\cal D}_k\}_{k\in\Omega}$ is such that    
 $a\in Dom({\cal D})$ then ${\cal D}\models {\cal S}$.

\noindent
(iii) There is an ${\cal S}'\in {\mathbb A}_s$ such that  ${\cal S}$ and ${\cal S}'$ are isomorphic.

\noindent
(iv) Suppose $f\in {\mathbb F}_s$ and $h$ is a homomorphism from $A[R(f)]$ (or $A[G(f)]$)  to $Chase({\cal T}_{\cal Q}, \spg{}{})$ such that $h(z)=a$. Then
there exists a homomorphism from $A[R(f)]$ (resp. $A[G(f)]$) to ${\cal S}'$.
}

The sense of Claim (iii) is that {\em a priori} $\cal S$ could have more than just two calves of the color that is opposite to the color of its head, and
then, still being a real s-pider, it would not be isomorphic to anything in ${\mathbb A}_s$. Proof of the Lemma (which we skip) is straightforward induction, using Lemma \ref{outdegree1}.

\lemmaaa{\label{domknietosc}
Let $zoo(\cal Q)$ be the set of all  s-piders ${\cal S}\in {\mathbb A}_s$ which are isomorphic to some  real s-pider in $Chase({\cal T}_{\cal Q}, \spg{}{})$. Then 
$zoo(\cal Q)$ is the smallest subset of ${\mathbb A}_s$ containing $\spg{}{}$ and closed under functions from $\cal Q$.
}

\noindent {\em Proof:} We know, from Lemma \ref{algebra} that if ${\cal S}'=f({\cal S})$ for some $f\in \cal Q$ and some  ${\cal S}\in {\mathbb A}_s$ then 
$Chase({\cal T}_{\cal Q}, \spg{}{})\models {\cal S}'$. To see that $zoo(\cal Q)$ is closed under functions from  $\cal Q$ notice that 
if $Chase({\cal T}_{\cal Q}, \spg{}{})\models {\cal S}$ then $Chase({\cal T}_{\cal Q}, {\cal S})$ is a substructure of $Chase({\cal T}_{\cal Q}, \spg{}{})$ and, in 
consequence, also $Chase({\cal T}_{\cal Q}, \spg{}{})\models {\cal S}'$.
For the opposite direction use Lemma \ref{1spider}.\eop

One more lemma we will need is:

\lemmaaa{\label{redlower}
Suppose each query in $\cal Q$ is lower. Then a s-pider in $zoo(\cal Q)$ is red if and only it is lower.
}

\noindent {\em Proof:} By usual induction on the fair sequence.

\subsection{Idempotence and sterile s-piders}\label{sterile}

 Notice that it very well may be the case that more than one copy of some ${\cal S}\in {\mathbb A}_s$ will be created in   $Chase({\cal T}_{\cal F},\spg{}{})$.

 Let for example ${\cal Q}$ be $\{\fff_1, \fff_2, \fff^3_1, \fff^3_2 \}$.  Then a copy of $\spg{3}{}$  can be constructed by first applying $\fff_1$ to $\spg{}{}$, 
and then $\fff^3_1$ to the resulting 
$\spr{}{1}$. But a different copy of   $\spg{3}{}$  will be produced by  first applying $\fff_2$ to $\spg{}{}$, and then $\fff^3_2$ to the resulting 
$\spr{}{2}$.

Imagine however that, after constructing (a copy ${\cal S}$ of) $\spg{}{3}$ in the first way, as $\fff^3_1(\fff_1(\spg{}{}))$
 we  try to apply $\fff^3_1$ to $\cal S$.
 According to Lemma \ref{algebra} it is of course possible,
and the result is a copy of $\spr{}{1}$. 
But it is not a new copy:  it follows from Lemma \ref{idempotencja} that  second consecutive use of  TGDs generated by the same
 query does not add to the Chase. 
In this context the following  lemma will be particularly useful:

\lemmaaa{[ 2-lame s-piders are sterile]\label{sterile1}
Suppose $\cal S$ is a real 2-lame  s-pider in some stage $\cal D$ of $Chase({\cal T}_{\cal Q}, \spg{}{})$. 
Then  $\cal S$ will never be used as a left hand side of a TGD execution leading to one of the later stages.
}

\noindent {\em Proof:} Suppose $\cal S$ is isomorphic to $\spr{i}{j}$ (the proof does not change if $\cal S$ is green).
It follows from Lemma \ref{algebra} that $\cal S$   could only be a result of applying the TGD 
$(\fff^i_j)^{G\rightarrow R}$ to $\spg{}{}$. But the only TGD that matches with $\cal S$ is $(\fff^i_j)^{R\rightarrow G}$.  
By Lemma \ref{idempotencja} it cannot however be now applied.\eop

Notice that, whenever we have $\cal Q$ containing 2-lame queries, like in  Subsection \ref{grosiagalnosc}, some
sterile 2-lame red s-piders will appear in $Chase({\cal T}_{\cal Q}, \spg{}{})$.

%%%%%%%%%%%%%%%%%%%%%%%%%%%%%%%%%%%%%%%%%%%%%%%%%%%%%%%%%%%%%%%%%%%%%%%%%%%%%%%%%%%%%%%%%%%%%%%%% BINARNE %%%%%
%%%%%%%%%%%%%%%%%%%%%%%%%%%%%%%%%%%%%%%%%%%%%%%%%%%%%%%%%%%%%%%%%%%%%%%%%%%%%%%%%%%%%%%%%%%%%%%%% BINARNE

\outline{
The queries in ${\mathbb F}_s$ (or functions, depending on what level of abstraction one wants to see them) which we considered so far 
were unary, in the sense that 
 they acted on single s-piders. In the rest of the paper we want them to 
be binary, so that they can {\bf rewrite words from ${\mathbb A}^*_s$ in a context-sensitive way}. And the ability to encode 
such a rewriting is a key to undecidability.}
 
%%%%%%%%%%%%%%%%%%%%%%%%%%%%%%%%%%%%%%%%%%%%%%%%%%%%%%%%%%%%%%%%%%%%%%%%%%%%%%%%%%%%%%%%%%%%%%%%% BINARNE %%%%%
%%%%%%%%%%%%%%%%%%%%%%%%%%%%%%%%%%%%%%%%%%%%%%%%%%%%%%%%%%%%%%%%%%%%%%%%%%%%%%%%%%%%%%%%%%%%%%%%% BINARNE %%%%%%%%%%%%%%%%%%%%%%%%%%%%%%%%%%%%%%%%%%%%%%%%%%%%%%%%%%%%%%%%%%%%%%%%%%%%%%
%%%%%%%%%%%%%%%%%%%%%%%%%%%%%%%%%%%%%%%%%%%%%%%%%%%%%%%%%%%%%%%%%%%%%%%%%%%%%%%%%%%%%%%%%%%%%%%%% BINARNE %%%%%

\section{Binary queries}\label{binarne}

We will now define two operations --  $\oaa$ and $\obb$ -- each of them taking two queries from  ${\mathbb F}_s$ and returning new ''binary'' 
query,
from the set that we will call ${\mathbb F}_s^2$.

It is maybe  good to recall here what are the free variables of the s-pider queries from ${\mathbb F}_s$: 2-lame queries have two free variables, and 
1-lame queries have one: the free variables are the knees of the legs with missing calves. When a s-pider query $f$ 
is seen as a green-red TGD, the free variables are what connects the new part of the structure, added by a single execution of a TGD, to the the old 
part\footnote{They of course also connect via the constants.}.

\definitionnn{
For $f, f'\in{\mathbb F}_s$ consider disjoint copies ${\mathbb G}$ of $A[f]$ and ${\mathbb G}'$ of $A[f']$. Let $V$ and $V'$ be  the sets of elements of ${\mathbb G}$ and ${\mathbb G}'$ and let 
 $W$ and $W'$ be the subsets of $V$ and $V'$ consisting of free variables of $f$ and $f'$. Let $z_2$ and  $z_1$ be the antenna and tail of $f$ and let 
   $z'_2$ and  $z'_1$ be the antenna and tail of $f$. Let $U(f,f')$ be the (disjoint) union of ${\mathbb G}$ and ${\mathbb G}'$. Then:

\noindent
\textbullet~ $f\oaa f'$ is the unique conjunctive query whose canonical structure is  $U(f,f')$, with $z_2$ and $z'_2$ identified, and with the set of free variables
equal to $W\cup W'\cup \{z_1,z'_1\}$;

\noindent
\textbullet~ $f\obb f'$ is the unique conjunctive query whose canonical structure is  $U(f,f')$, with $z_1$ and $z'_1$ identified, and with the set of free variables
equal to $W\cup W'\cup \{z_2,z'_2\}$.
}

The set of all  $f\oaa f'$  (or $f\obb f'$) for $f,f'\in {\mathbb F}_s$ will be called ${\mathbb F}_s^\oaa$ (resp. ${\mathbb F}_s^\obb$). 
We also define ${\mathbb F}_s^2$ as ${\mathbb F}_s^\oaa \cup {\mathbb F}_s^\obb$.

The main lemma, obviously implying Theorem \ref{main} is:

\lemmaaa{\label{main1}
It is an undecidable problem to determine, for given $s\in \mathbb N$ and given 
 $Q\subseteq {\mathbb F}_s^2$ whether Chase(${\cal T}_{\cal Q}, \spg{}{})\models \spr{}{}$.
}

The rest of the paper is devoted to the proof of this lemma.

\section{Abstracting from the s-pider details}

Let $Q\subseteq {\mathbb F}_s^2$ be a  set of binary queries. We would like to understand the structure of
 Chase(${\cal T}_{\cal Q}, \spg{}{}$) so that, in particular, we understand when Chase(${\cal T}_{\cal Q}, \spg{}{})\models \spr{}{}$ holds.

First of all notice that Lemma \ref{outdegree1} and \ref{1spider} 
survive in the new context -- together with their proofs. 
But be careful here:

\lemmaaa{\label{1spiderc}
For each pair $a_1,a_2 $ of elements of Chase(${\cal T}_{\cal Q}, \spg{}{})$ there are at most two elements $a$ 
 such that Chase(${\cal T}_{\cal Q}, \spg{}{})\models ${\small H}$(a, a_1,a_2)$. 
}

\noindent {\em Proof:} Induction. For the induction step notice that an atom  {\small H}$(a,a_1,a_2)$ can only be created 
together with either a new element $a_1$ (if a TGD generated by a query from ${\mathbb F}^\obb_s$  is used) or with a new 
 $a_2$ (when the query is from ${\mathbb F}^\oaa_s$). And notice that
 a single execution of a TGD generated by a  query from  ${\mathbb F}_s^2$ creates two atoms of the predicate {\small H}, and 
the newly created $a_1$ occurs in both of them. Notice that the  newly created spiders are always both of the same color.\eop

\subsection{Queries $f_1\oaa f_2$ and $f_1\obb f_2$ in action}\label{tralalala} 

Let now $Q\in \cal Q$ be of the form\footnote{The case when $Q$ is of the form $f_1\obb f_2$ is analogous.} $f_1\oaa f_2$ and suppose ${\mathbb D}$ is a structure (a stage of Chase(${\cal T}_{\cal Q}, \spg{}{}$)).

Consider  the TGD $Q^{R\rightarrow G}$ and let us try to imagine how this TGD could be executed in ${\mathbb D}$.
First  a homomorphism $h$ from $A[R(Q)]$ to ${\mathbb D}$ needs to  be found. 

$A[R(Q)]$ contains 3 antenna/tail vertices: $ z_1, z_1'$ and $z_2$, joined by the atoms $R(${\small H}$)(z,z_1, z_2)$ and  
$R(${\small H}$)(z', z_1', z_2)$.
This means that two  red atoms $R(${\small H}$)(h(z), h(z_1), h(z_2))$ and  $R(${\small H}$)(h(z'), h(z_1'), h(z_2))$
 must be found in  ${\mathbb D}$. 

Notice that, due to Lemma \ref{1spiderc}, once $h(z_1)$, $h(z_1')$ and $h(z_2)$     
are fixed
there are at most two possible choices for each of $h(z)$ and $h(z')$. And   once $h(z)$ and $h(z')$ are fixed then, due to Lemma \ref{1spider}
there is exactly 
one real s-pider  ${\cal S}_1$ in ${\mathbb D}$ with $R(${\small H}$)(h(z), h(z_1), h(z_2))$ 
and exactly one
real s-pider  ${\cal S}_2$ in ${\mathbb D}$ with $R(${\small H}$)(h(z'), h(z_1'), h(z_2))$ 

Now, in order for the query $A[R(Q)]$ to be executed we need  the query $R(f_1)$ to match with   ${\cal S}_1$  and 
 the query $R(f_2)$ to match
with   ${\cal S}_2$.
Lemma \ref{algebra} tells us when it is possible.

Once the triples $h(z), h(z_1), h(z_2)$  and  $h(z'), h(z_1'), h(z_2)$, satisfying all the above constraints, are found,
a copy  of $A[G(Q)]$  is created\footnote{Unless it already existed.}, consisting of 
two  green\footnote{Of course this is all true also for the colors reversed.} s-piders $f_1({\cal S}_1)$ and $f_2({\cal S}_2)$. 
This is because on the level of individual s-piders we are exactly in the world of Section \ref{monadyczne}.

What is new is how the two new s-piders 
%(that we imagine just as green $P$-edges labeled with $f_1({\cal S}_1)$ and $f_2({\cal S}_2)$) 
are connected to each other and to
the old part of the structure: the antenna of $f_1({\cal S}_1)$ is a new element  -- it was quantified in $Q$ -- 
and is identified with the antenna of $f_2({\cal S}_2)$.
 
But the tail of $f_1({\cal S}_1)$ was free in $Q$ so it is identified with  $h(z_1)$ and the tail of  $f_2({\cal S}_2)$ was free and it is
identified with $h(z_2')$. So the new copy of $A[G(Q)]$ is connected to the old structure via the tails of the two new s-piders.

Of course the two new s-piders are also connected to the old structure via the free variables (and constants) which are not in their $H$ atoms. 
But there are two reasons why we do not need to bother about it. First of them is Lemma \ref{1spider}. Second is that, while each TGD generated by a query from ${\mathbb F}^2_s$
needs two spiders to be executed, and requires them to share their antennas (or tails), it is oblivious to any other possible connections between the two 
s-piders (via knees).  
This analysis shows that we now can completely abstract from the low-level implementation details of the s-piders,
in particular from details like the relations $C^i$, $C_i$, $T^i$, $T_i$ and concentrate on the high-level notions.

\subsection{S-warm and s-warm rewriting rules} 

A s-warm is defined as a multi-labeled graph (which means that each edge can have one or more labels),
 whose edges are the {\small H} atoms of some structure (intended to be a stage
 of Chase(${\cal T}_{\cal Q}, \spg{}{})$):

\definitionnn{
A s-warm $\mathbb D$ is a ternary relation {\small H}$\subseteq {\mathbb A}_s\times D\times D$. To keep notations light we 
use the term ''elements of  $\mathbb D$'' for elements of $D$. Elements of  ${\mathbb A}_s$ are {\em labels}. 
We assume that for each two elements $a,b$ of a s-warm there are at most two s-piders $\cal S$  such that {\small H}$({\cal S},a,b)$,
and that they are of the same color. 
 Atoms {\small H}$({\cal S},a,b)$, or just pairs  $a,b$,  such that ${\mathbb D}\models ${\small H}$({\cal S},a,b)$, for some $\cal S$, are called {\em edges}. An edge is green or red,
depending on the  s-piders being its labels.
}

We are going to see queries from ${\mathbb F}^2_s$ as s-warm rewriting rules:

\definitionnn{
Let $Q= f\oaa f'$ (or $Q= f\obb f'$) be  from ${\mathbb F}^2_s$ and let $\mathbb D$ be a s-warm. We say that 
a rewriting $Q$ can be executed in 
$\mathbb D$ if:

\ding{184} there are edges {\small H}$({\cal S},a,b)$ and {\small H}$({\cal S}',a',b)$ (resp. {\small H}$({\cal S},a,b)$ and {\small H}$({\cal S}',a,b')$), 
 such that ${\cal S}\in Dom(f)$ and ${\cal S}'\in Dom(f')$ are both of the same color;

\ding{185} there is  no $b'$ (resp. $a'$) such that {\small H}$(f({\cal S}),a,b')$ and {\small H}$(f'({\cal S}'),a',b')$ 
  (resp. {\small H}$(f({\cal S}),a',b)$ and {\small H}$(f'({\cal S}'),a',b')$) are edges of $\mathbb D$.

Pair of edges {\small H}$({\cal S},a,b)$ and {\small H}$({\cal S}',a',b)$ is called {\em the input of the rewriting} (notice that order is important here).
The result  of the rewriting  is then a new structure ${\mathbb D}'$ being $\mathbb D$ with new vertex $b'$ (resp. $a'$) and 
new edges {\small H}$(f({\cal S}),a,b')$ and {\small H}$(f'({\cal S}'),a',b')$  (resp.  {\small H}$(f({\cal S}),a',b)$ and {\small H}$(f'({\cal S}'),a',b')$) as above.
}

Notice that we did not require in the above definition that $a\neq a'$ (resp.  $b\neq b'$). Not only we have no means to enforce such requirement, but also, 
 since we begin the Chase from a single (full green) s-pider  the possibility of having them equal is of crucial importance for us.

See that -- while we are not literally talking about TGDs now -- conditions \ding{184} and \ding{185} are analogous to  \ding{182} 
and \ding{183} from Section \ref{tgds} and we can still (like in Section \ref{chase}) 
define a fair (with respect to a set $\cal Q$ of rewritings and an original s-warm ${\mathbb D}$) 
sequence of structures $\{{\cal C}_k\}_{k\in\Omega}$, with each ${\cal C}_{k}$ 
being a result of  a single execution of a rewriting rule in the structure   $\bigcup_{l<k}{\cal C}_{k}$ and with each possible rewriting ultimately being executed. 
We can also define the fixpoint of the rewritings, as the 
union of $\bigcup_{k\in \Omega}{\cal C}_k$. 
To distinguish, we will call the union $chase({\cal Q}, {\mathbb D})$. 

\subsection{The abstraction}

Let $\mathbb D$ be a structure over $\Sigma$, such that if $\mathbb D\models ${\small H}$(a,b,c)$ then there is exactly
 one real s-pider $\cal S$ in $\mathbb D$ having $a$ as its head,
and such that each real s-pider in $\mathbb D$ is isomorphic to some  element of $\mathbb A$. The following definition
and lemma hardly come as a surprise:

\definitionnn{ The s-warm
s-warm$(\mathbb D)$ is defined as the
set of all  triples  {\small H}$({\cal S},b,c)$  such that
$\mathbb D\models ${\small H}$(a,b,c)$ and $a$ is the head of a real s-pider in $\mathbb D$ which  is isomorphic to $\cal S$.
}

From now on let   ${\mathbb D}_{\spg{}{}}$ be the s-warm consisting of a single  edge labeled with $\spg{}{}$. 
Define $\cal F$ as the set of all fair (with respect to the set  ${\cal T}_{\cal Q}$ of TGDs and the structure $\spg{}{}$) 
 sequences $\{{\cal D}_k\}_{k\in \Omega}$ and let 
 ${\cal F}'$ be the set of all fair (with respect to the set ${\cal Q}$ of rewriting rules and the s-warm ${\mathbb D}_{\spg{}{}}$) 
 sequences $\{{\cal C}_k\}_{k\in \Omega}$.

\lemmaaa{\label{abstrakcja}
The mapping that maps a sequence $ \{ {\cal D}_k \}_{k\in \Omega}$ 
of structures to a sequence 
$\{$s-warm$({\cal D}_k)\}_{k\in \Omega}$
of s-warms
 is a bijection from $\cal F$ to  ${\cal F}'$.
}

For a given set
$Q\subseteq {\mathbb F}_s^2$ let ${\cal C}^{\cal Q} = chase({\cal Q}, {\mathbb D}_{\spg{}{}})$.
From now on we forget about  $Chase({\cal T}_{\cal Q}, \spg{}{})$  and TGDs and concentrate on s-warms and their rewritings.
Due to Lemma \ref{abstrakcja}, in order to prove Lemma \ref{main1} it is enough to show: 

\lemmaaa{\label{main2}
It is an undecidable problem to determine, for given $s\in \mathbb N$ and given set
 $Q\subseteq {\mathbb F}_s^2$ of rewriting rules, whether ${\cal C}^{\cal Q}$
 contains any edge labeled with $\spr{}{}$.
}

\subsection{One more lemma}

Before we end this Section it will be maybe illuminating to notice one peculiar property of  
${\cal C}^{\cal Q}$. The proof of the following lemma goes by easy induction:

\lemmaaa{Let $Q\subseteq {\mathbb F}_s^2$. Then
each vertex of ${\cal C}^{\cal Q}$ either has in-degree zero (such vertex  will be called tail, as  
 it is the tail of all the edges it belongs to)
or has out-degree zero (and it is the antenna of all the edges it belongs to). This implies that all the directed $H$-paths in 
 ${\cal C}^{\cal Q}$ have length one. 
}

Now imagine vertices of ${\cal C}^{\cal Q}$  drawn in two rows -- all the antennas in the upper row and all the tails in the lower one -- and
see how mnemonic the fonts $\oaa$ and $\obb$ are.

\section{An important example (quite complicated)}

Consider a set $\cal Q_{\eta}$ consisting of the following three pairs of {\em associated} rewritings:

\ding{192}~~ {\bf A:} $\fff_1  \oaa \fff_2 $~ ~and~   {\bf B:} $\fff^\alpha_1  \oaa \fff^{\eta_1}_2 $\smallskip 

\ding{193}~~ {\bf A:} $\fff^{\eta_0}_3  \oaa \fff_4 $ ~and~  {\bf B:} $\fff^{\beta_0}_3  \oaa \fff^{\eta_1}_4 $ \smallskip 

\ding{194}~~ {\bf A:} $\fff^{\eta_1}_5  \obb \fff_6 $   ~and~ {\bf B:} $\fff^{\beta_1}_5  \obb  \fff^{\eta_0}_6 $\smallskip

where $\alpha , {\beta_0}, {\eta_0}, {\beta_1}, {\eta_1}\in \mathbb S$,

And let us try to have a glimpse of ${\cal C}^{{\cal Q}_{\eta}}$.
Let {\small H}$(\spg{}{},s_0,t_0)$ be the only edge of ${\mathbb D}_{\spg{}{}}$.

%  First of all notice that we can execute an dishonest rewriting, using rule \ding{192} {\bf A} and  {\small H}$(s,t)$ as both the inputs. A new vertex
% $t'$ will be created, together with edges {\small H}$(s,t')$ and {\small H}$(t',s)$, labelled with $\spr{}{1}$ and   $\spr{}{2}$. 

The table  describes a (finite prefix of) an infinite sequence of rewritings that will be of special importance for us.
Newly created elements are marked with bold.\medskip

\hspace{-6mm}
{\small
\begin{tabular}{ | l |l | l | }
\hline

        Input           & \hspace{-1.5mm}Rule           \hspace{-1mm}            &   Output                                              \\
        edges           & \hspace{-1.5mm}used            \hspace{-1mm}           &   edges                                                 \\\hline
 \hspace{-3mm} {\small H}$(\spg{}{},s_0,t_0)$,  {\small H}$(\spg{}{},s_0,t_0)$  &\hspace{-1.5mm}\ding{192}A\hspace{-1mm}&\hspace{-1.5mm}{\small H}$(\spr{}{1},s_0,{\bf t'})$, {\small H}$(\spr{}{2},s_0, {\bf t'})$  \\                                  
 \hspace{-3mm} {\small H}$(\spr{}{1},s_0,t')$, {\small H}$(\spr{}{2},s_0, t')$ &\hspace{-1.5mm}\ding{192}B\hspace{-1mm}&\hspace{-1.5mm}{\small H}$(\spg{\alpha}{},s_0,{\bf t_1})$, {\small H}$(\spg{\eta_1}{},s_0, {\bf t_1})$ \\                                     
 \hspace{-3mm} {\small H}$(\spg{\eta_1}{},s_0, t_1)$, {\small H}$(\spg{}{},s_0,t_0)$ &\hspace{-1.5mm}\ding{193}A\hspace{-1mm}&\hspace{-1.5mm}{\small H}$(\spr{}{5},{\bf s'},t_1)$, {\small H}$(\spr{}{6},{\bf s'},t_0)$ \\                                         
  \hspace{-3mm} {\small H}$(\spr{}{5},s',t_1)$, {\small H}$(\spr{}{6},s',t_0)$  &\hspace{-1.5mm}\ding{193}B\hspace{-1mm}&\hspace{-1.5mm}{\small H}$(\spg{\beta_1}{},{\bf s_1}, t_1)$, {\small H}$(\spg{\eta_0}{},{\bf s_1}, t_0)$                                          \\  
  \hspace{-3mm} {\small H}$(\spg{\eta_0}{},{\bf s_1}, t_0)$, {\small H}$(\spg{}{},s_0,t_0)$   &\hspace{-1.5mm}\ding{194}A\hspace{-1mm}&\hspace{-1.5mm}{\small H}$(\spr{}{3}, s_1,{\bf t''})$, {\small H}$(\spr{}{4},s_0,{\bf t''})$                                            \\
 \hspace{-3mm} {\small H}$(\spr{}{3}, s_1, t'')$, {\small H}$(\spr{}{4},s_0, t'')$  &\hspace{-1.5mm}\ding{194}B\hspace{-1mm}&\hspace{-1.5mm}{\small H}$(\spg{\beta_0}{}, s_1,{\bf t_2})$, {\small H}$(\spg{\eta_1}{},s_0,{\bf t_2})$ \\
 \hspace{-3mm} {\small H}$(\spg{\eta_1}{},s_0, t_2)$, {\small H}$(\spg{}{},s_0,t_0)$ &\hspace{-1.5mm}\ding{193}A\hspace{-1mm}&\hspace{-1.5mm}{\small H}$(\spr{}{5},{\bf s''},t_2)$, {\small H}$(\spr{}{6},{\bf s''},t_0)$ \\\hline

\end{tabular}\\\medskip  
}

Compare the two rewritings using the rule \ding{193}A and notice the recursion. Then proving the following lemma will be an easy exercise:

\lemmaaa{\label{tasma}
There are infinite sequences $t_1,t_2, \ldots $ and $s_1, s_2, \ldots $ of elements  of ${\cal C}^{{\cal Q}_{\eta}}$
such that:
 {\small H}$(\spg{\alpha}{},s_0,t_1)$, and for each $k$ there is {\small H}$(\spg{\beta_1}{},s_k,t_k)$ and 
{\small H}$(\spg{\beta_0}{},s_k,t_{k+1})$ and  {\small H}$(\spg{\eta_1}{},s_0,t_k)$.
}

\section{Friendly Thue systems}

We will now consider  Thue systems $\Pi\subseteq {\mathbb S}^* \times {\mathbb S}^*$. Elements of  ${\mathbb S}$ are numbers, 
so some of them are even and some are odd. We think that  $\alpha $,  ${\beta_0}$ and  ${\eta_0}$ are even and  ${\beta_1}$ and  $\eta_1$ are odd. 

A set of productions of a Thue system $\Pi\subseteq {\mathbb S}^* \times {\mathbb S}^*$ will be called {\em friendly} if
$\Pi = \Pi_<\cup \Pi_=$ where:

\noindent\textbullet~
$\Pi_<$ consists of two pairs
$ \{ \eta_0, \beta_0\eta_1 \}$ and  $\{ \eta_1, \beta_1\eta_0 \}$;

\noindent\textbullet~all  productions of $\Pi_=$ are of the form   $\{ij,i'j'\}$ for $i,i',j,j'\in {\mathbb S}$;

\noindent\textbullet~ if $\{ij,i'j'\}\in\Pi_=$ then both $i$ and $i'$ are odd and  both $j$ and $j'$ are even or  both $i$ and $i'$ are even and  both $j$ and $j'$ are odd;

\noindent\textbullet~ there is no production of $\Pi_=$ of the form  $\{ij, ij'\}$ or  $\{ij,i'j\}$;

\noindent\textbullet~ there is no production involving $\alpha$; no production in $\Pi_=$ involves $\eta_0$ or $\eta_1$;

\noindent\textbullet~ there is an odd $\gamma\in {\mathbb S}$, and even $\gamma'\in {\mathbb S}$ which occur (each of them) in 
 exactly one production of $\Pi$, which is $\{ii',\gamma\gamma'\}$ for some $i,i'\in {\mathbb S}$;

\noindent\textbullet~ $s> 2|\Pi|$ 

It is easy to prove (using the techniques presented in \cite{D77}) that the problem:

\shaded{For a set of productions of a friendly Thue system $\Pi$, do there exist ${\bf w,w'}\in  {\mathbb S}^*$ such that  
${\bf w}\gamma\gamma'{\bf w'}\stackrel{\ast\;\;\;}{\Leftrightarrow_\Pi}\alpha\eta_1$\hfill ?
}

is undecidable.

Let now  $\Pi$ be a fixed  friendly Thue system.  Lemma \ref{main2}, and therefore Theorem \ref{main}, will be proved once we 
 construct a set $\cal Q$ of rewritings such that the two conditions are equivalent:\smallskip

\noindent
\smiley: There is an edge, in ${\cal C}^{\cal Q}$ labeled with  $\spr{}{}$;

\noindent
\blacksmiley:~ ${\bf w}\gamma\gamma'{\bf w'}\stackrel{\ast\;\;\;}{\Leftrightarrow_\Pi}  \alpha\eta_1$ for some ${\bf w,w'}\in  {\mathbb S}^*$.\smallskip

The following Lemma is easy to prove and will be useful:

\lemmaaa{\label{najpierwdlugosc}
Condition \blacksmiley ~holds if and only if
there is  $m\in \mathbb N$ such that 
${\bf w}\gamma\gamma'{\bf w'}\stackrel{\ast\;\;\;\;\;}{\Leftrightarrow_{\Pi_=}}  \alpha(\beta_1\beta_0)^m\eta_1$ for some ${\bf w,w'}\in  {\mathbb S}^*$.

}

\subsection{The set $\cal Q$}

 First we define ${\cal Q}_0$, as the  set of rewritings  consisting of:

\noindent-- all the rewriting rules from  $\cal Q_{\eta}$

\noindent--   two {\em associated} rewriting rules  
\ding{195} $\fff^i_{l_p}\oaa \fff^j_{r_p}$ and $\fff^{i'}_{l_p}\oaa \fff^{j'}_{r_p}$ for each production  $p=\{ij, i'j'\}$ in $\Pi$ with $i$ even;

\noindent--   two {\em associated} rewriting rules  
\ding{196} $\fff^i_{l_p}\obb \fff^j_{r_p}$ and $\fff^{i'}_{l_p}\obb \fff^{j'}_{r_p}$ for each production  $p=\{ij, i'j'\}$ in $\Pi$ with $i$ odd;

where each of the numbers $l_p$ and each $r_p$ only occurs in the two aforementioned rewriting rules. 
Finally, $\cal Q$ is defined as $Q_0$ with one additional rewriting rule
\noindent\ding{197}  $\fff^\gamma\obb \fff^{\gamma'}_{r}$
with $r$ not occurring anywhere in the rules of ${\cal Q}_0$.

\outline{\noindent The rest of the paper is devoted to understanding  the structure, first  of ${\cal C}^{{\cal Q}_0} = chase({\cal Q}_0,{\mathbb D}_{\spg{}{}})$ 
 and then of   ${\cal C}^{{\cal Q}} =chase({\cal Q},{\mathbb D}_{\spg{}{}})$
in order to prove that  \smiley~  holds true if and only of \blacksmiley~ does.\ In the following Section \ref{hunt} we prove that condition
\blacksmiley~ implies \smiley.}

\section{How to hunt a (full) red s-pider}\label{hunt}

\definitionnn{\label{okulary} For a s-warm  $\mathbb D$
the set $\bar W({\mathbb D}) \subseteq {\mathbb S}^*\times {\mathbb D}^2$ is defined as the smallest set such that:\smallskip

\noindent
\textbullet~$\langle \varepsilon , a, a \rangle\in \bar W({\mathbb D}) $ for each $a\in{\mathbb D}$;

\noindent
\textbullet~if $\langle w, a, b\rangle \in \bar W({\mathbb D}) $ and 
${\mathbb D}\models {\small H}(\spg{i}{},b,b')$ for some even $i$ then $\langle wi, a, b' \rangle\in \bar W({\mathbb D}) $;

\noindent
\textbullet~if $\langle w, a, b \rangle \in \bar W({\mathbb D}) $ and 
${\mathbb D}\models ${\small H}$(\spg{i}{},b',b)$ 
for some odd $i$ then $\langle wi, a,b' \rangle\in \bar W({\mathbb D}) $.

\noindent
Then  $W({\mathbb D})\subseteq {\mathbb S}^*$ is defined as $\{w: \exists a, b\; \langle w, a, b \rangle\in \bar W({\mathbb D})\}$.
}

In other words $W({\mathbb D})$ is the set of all words that can be constructed as follows: walk an undirected  path in ${\mathbb D}$
(form some $a$ to some $b$)
 and read (and remember)  the labels of all the edges you cross.   But you are only
allowed to take  edges labeled by green 1-lame s-piders. And if the label is $\spg{i}{}$, for an even $i$, 
then you must walk in the direction of the edge,
otherwise you must walk in the opposite direction.

\lemmaaa{\label{thuedziala}
If ${\bf v}\in W({\cal C}^{{\cal Q}_0})$ and ${\bf v} \stackrel{\ast\;\;\;\;\;}{\Leftrightarrow_{\Pi_=}} {\bf v'}$ then also   
${\bf v'}\in W({\cal C}^{{\cal Q}_0}) $.
}

\noindent {\em Proof:} By induction it is enough to prove that if ${\bf v}\in W({\cal C}^{{\cal Q}_0}) $ 
and ${\bf v} \stackrel{\;\;\;\;\;}{\Leftrightarrow_{\Pi_=}} {\bf v'}$ then also
   ${\bf v'}\in W({\cal C}^{{\cal Q}_0}) $.
Suppose that ${\bf v}={\bf w_1}ij{\bf w_2}$, that   ${\bf v'}={\bf w_1}i'j'{\bf w_2}$ with $\{ij,i'j'\}\in \Pi_=$ and $i$ even (the other case is analogous) and that ${\bf v}\in W({\cal C}^{{\cal Q}_0}) $. 
The last assumption means that there are 
vertices $a,b,c,d,e$ in ${\cal C}^{{\cal Q}_0}$ such that both the triples $\langle {\bf w_1},a,b\rangle$ and  
$\langle {\bf w_2},d,e\rangle$ are in $\bar W({\cal C}^{{\cal Q}_0})$ and
edges 
{\small H}$(\spg{i}{},b,c)$ and 
{\small H}$(\spg{j}{},d,c)$ are in ${\cal C}^{{\cal Q}_0}$. 

By the assumption that $\{ij,i'j'\}\in \Pi_=$ and $i$ even we have that the rewritings 
$\fff^i_{l_p}\oaa \fff^j_{r_p}$ and $\fff^{i'}_{l_p}\oaa \fff^{j'}_{r_p}$ are in $\cal Q$. But -- since ${\cal C}^{{\cal Q}_0}$ is
closed under rewritings -- the  first of these rewritings enforces that there 
must be a $c'$ in ${\cal C}^{{\cal Q}_0}$ with 
{\small H}$(\spr{}{l_p},b,c')$ and 
{\small H}$(\spr{}{r_p},d,c')$. And the second of the rewritings enforces that there 
must be a $c''$ in ${\cal C}^{{\cal Q}_0}$ with 
{\small H}$(\spg{i'}{},b,c'')$ and 
{\small H}$(\spg{j'}{},d,c'')$. So also ${\bf v'}\in W({\cal C}^{{\cal Q}_0}) $.
\eop

\lemmaaa{\label{usmieszki}
Condition \blacksmiley~implies \smiley. 
}

\noindent {\em Proof:} 
By  Lemma \ref{najpierwdlugosc} condition \blacksmiley~implies that
there is  $m\in \mathbb N$ such that 
${\bf w}\gamma\gamma'{\bf w'} \stackrel{\ast\;\;\;\;\;}{\Leftrightarrow_{\Pi_=}}  \alpha(\beta_1\beta_0)^m\eta_1$ for some ${\bf w,w'}\in  {\mathbb S}^*$.

It follows from Lemma \ref{tasma}, and from Definition \ref{okulary} that  $\alpha(\beta_1\beta_0)^m\eta_1\in W({\cal C}^{{\cal Q}_0})$ for some  $m\in \mathbb N$. 
By Lemma \ref{thuedziala} this implies that the word $\gamma\gamma'$ is in $W({\cal C}^{{\cal Q}_0})$, 
which means that there are vertices $a,a',b$ of ${\cal C}^{{\cal Q}_0}$ such that 
{\small H}$(\spg{\gamma}{},a,b)$ and {\small H}$(\spg{\gamma'}{},a,b')$ hold in ${\cal C}^{{\cal Q}_0}$. Now use the rule $\fff^\gamma\obb \fff^{\gamma'}_{r}$ to produce an edge labeled with $\spr{}{}$.\eop

%%%%%%%%%%%%%%%%%%%%%%%%%%%%%%%%%%%%%%%%%%%%%%%%%%%%%%%%%%%%%%%%%%%%%%%%%%%%%%5

\outline{\noindent Now we only need to prove that condition \smiley~implies \blacksmiley. It is much more complicated than the opposite implication.

In the rest of the paper {\bf we assume that \blacksmiley~does not hold true}. Our goal is to show that
\smiley~is not true either. 
The plan is to first consider a sequence $\{{\cal C}_i\}_{i\in \mathbb \omega}$ (where $\omega$ is the first infinite ordinal) 
fair (with respect to ${\cal Q}_0$ and ${\mathbb D}_{\spg{}{}}$)  and analyze the structure  ${\cal C}^{{\cal Q}_0}=\bigcup_{i\in \omega}{\cal C}_i$. This will be done in Sections \ref{latwe}--\ref{bezczerwone}.

Then we must of course face the possibility that 
the list ToDo(\ding{197},${\cal C}^{{\cal Q}_0}$) will be very much non-empty and many (infinitely many) further
rewritings may be needed. But -- as we are going to prove  in Section \ref{ostatnia} -- 
all the edges created by these rewritings will be sterile.}

\section{Getting rid of the reds}\label{latwe}

First of all notice that all the rewritings used in ${\cal Q}_0$ are lower\footnote{It is not true about $\cal Q$ and this is the reason why we analyze 
${\cal C}^{{\cal Q}_0}$ first.}. The proof of the following lemma is by (easy) induction, almost the same as the proof of Lemma \ref{redlower}:

\lemmaaa{\label{czerwonedolne}
Let  ${\cal S}$ be  the label of some edge in ${\cal C}^{{\cal Q}_0}$. Then  ${\cal S}$ is red if and only if it is lower. In particular  $\spr{}{}$ is not a label of any edge in ${\cal C}^{{\cal Q}_0}$. Also $\spr{}{r}$ cannot be a label of any edge in ${\cal C}^{{\cal Q}_0}$ (where  $r$ is from rule \ding{197}).
}

\definitionnn{
Two red edges {\small H}$({\cal S},a,b)$  and {\small H}$({\cal S}',a',b)$ (or {\small H}$({\cal S},a,b)$  and {\small H}$({\cal S}',a,b')$) of ${\cal C}^{{\cal Q}_0}$ will be called {\em a married couple} if they were 
created in the same rewriting step. The vertex $b$ (resp. $a$) is called a {\em knot} then.
}

As it turns out, a knot is never touched by any edge other than the two spouses it joins:

\lemmaaa{\label{wezel} If $a$ is a knot then it has degree 2 in ${\cal C}^{{\cal Q}_0}$.
}

\noindent {\em Proof:} Suppose the knot $b$ was created, together with red edges {\small H}$({\cal S},a,b)$  and {\small H}$({\cal S}',a',b)$ 
by an execution of some rule $\fff^I_i\oaa \fff^J_j$ (the case with
$\fff^I_i\obb \fff^J_j$ is analogous). This rule was 
applied to two green edges with labels $\spg{I'}{}, \spg{J'}{}$  and thus ${\cal S}=\spr{I \setminus I'}{i} $ and 
 ${\cal S}'=\spr{J\setminus J'}{j}$. 

The only way to create a new edge containing the vertex $b$, 
would be to find some edge {\small H}$({\cal S}_1,a,b')$ (or  {\small H}$({\cal S}_1,a',b')$) and use a
rule of the form $f\obb f'$. But if $f$ occurs in any rule from ${\mathbb F}_s^\obb$ then it cannot be applied to any s-pider of the form $\spr{K}{i}$ -- this is because
of the assumption that each of  the lower subscripts can only occur in  two associated rewritings.\eop

\lemmaaa{[No children out of wedlock, whatever temptation]\label{wmalzenstwie}
Suppose {\small H}$({\cal S},a,b)$ is an element of a married couple of reds in some ${\cal C}_n$, created by    
an execution of some rule  $f$ from ${\cal Q}_0$. 
Then the only way for it to be a part of the 
input of any future rewriting by a rule $g$ from ${\cal Q}_0$
is
that the other element of the input of this rewriting is its spouse and that $g$ is the rule associated with $f$. 
}

\noindent{\em Proof:} Suppose $f\in {\mathbb F}_s^\obb$  (the other case is analogous), so $a$ is the knot joining {\small H}$({\cal S},a,b)$ with its spouse.
It follows from Lemma \ref{wezel} that  $g\in {\mathbb F}_s^\obb$ -- otherwise the degree of $a$ would be greater than 2 at some point. 
So the only way for an edge to be an input of a rewriting together with    {\small H}$({\cal S},a,b)$ is to contain the vertex $a$. But $a$ only belongs to two edges
in ${\cal C}^{{\cal Q}_0}$: to {\small H}$({\cal S},a,b)$ and its spouse. Using the argument from the proof of Lemma \ref{wezel}, that the numbers $l_p$ and  $r_p$ 
can only occur in  two associated rewritings, we get that $g$ is either $f$ itself (which is impossible due to idempotence) or is associated with $f$.\eop

Notice (and this remark will be needed in Section \ref{ostatnia}) 
that the proof does not rely on the shortage of possible candidates who would be keen to produce offspring with {\small H}$({\cal S},a,b)$. The reasons for 
its faithfulness is inherent to {\small H}$({\cal S},a,b)$ itself, and even someone like {\small H}$(\spr{}{},c,b)$ would not change its mind ($\spr{}{}$ being the most
promiscuous red label).

\lemmaaa{[Sterile reds]\label{sterilereds}
(i) If a red {\small H}$({\cal S},a,b)$ is an element of a married couple of red edges in some ${\cal C}_n$ and  ${\cal S}$ is 2-lame then 
neither {\small H}$({\cal S},a,b)$ nor its spouse are never used as an input of a rewriting rule execution.

\noindent(ii) A rewriting in which $\spg{}{}$ is used as an input of any 2-lame rewriting rule from 
${\mathbb F}_s$ leads to a pair of sterile red edges. 
}

 Proof of this Lemma is left as an easy exercise. Use the assumption that there is no production in $\Pi$ of the form  $\{ij, ij'\}$ or  $\{ij,i'j\}$
and the argument from the proof of Lemma \ref{sterile1}.

From now on we assume that the sequence $\{{\cal C}_k\}_{k\in \mathbb N}$ is such, that whenever 
a married couple of reds  is created at some step, at the next step the only rewriting this red marriage is able to be the input of is executed (unless this married couple is sterile). So  we can imagine that we always execute procedures consisting of two associated rules, and produce greens from other greens.
The red edges are in the structure but in no way they contribute to its complexity and we do not need to think of them any more. 

\section{Dangerous vertices}\label{niebezpieczne}

In our quest to understand the structure of ${\cal C}^{Q_0}$ we now concentrate on the green edges.
We already know (from Lemma \ref{sterilereds}) that  2-lame rewritings applied to $\spg{}{}$ never produce anything relevant. 
Notice also that all the rewritings used in ${\cal Q}_0$ are lower, and 
all  green edges of ${\cal C}^{Q_0}$, apart from edges labeled with $\spg{}{}$, are upper.
This means that only 2-lame rewriting rules can be applied to edges with labels of the form $\spg{i}{}$. 

In particular this means that any  rewriting with the rule \ding{192}A must take as its input two edges 
labeled with $\spg{}{}$, rewritings \ding{193}A and \ding{194}A take one edge labeled
with $\spg{}{}$ and one with $\spg{i}{}$ and so on.

\definitionnn{
A vertex of $\cal C$, or any ${\cal C}_i$, is called {\em dangerous} if it is a tail or an antenna of some edge labeled with $\spg{}{}$.
}

\lemmaaa{\label{niebezpieczne}
Let {\small H}$(\spg{i}{},a,b)$ be a green edge of  ${\cal C}_i$. Then (i)
 $a$ is dangerous if and only if $i$ is either $\alpha$ or $\eta_0$
and (ii) $b$ is dangerous if and only if $i$ is $\eta_1$.
}

\noindent {\em Proof:} By induction on $i$. The claim is clearly true in ${\cal C}_0$ as it consists of a single edge labeled with $\spg{}{}$.

Suppose the claim  is true in some ${\cal C}_n$. Suppose the structure ${\cal C}_{n+2}$ is a result of first applying some rewriting $f$ to green edges 
 {\small H}$({\cal S}_1,a,b)$ and {\small H}$({\cal S}_2,a',b)$, 
(or to {\small H}$({\cal S}_1,a,b)$ and {\small H}$({\cal S}_2,a,b')$ -- -- in cases where rewriting rules \ding{194} or \ding{196} were used), creating 
two new red edges, and then  applying $f'$, associated with $f$, to the two new red edges (we know, from Section \ref{latwe}, that this is the only scenario
one needs to consider).

As a result a new vertex $b'$ (resp. $a'$) is created, together with new green edges 
{\small H}$({\cal S}'_1,a,b')$ and {\small H}$({\cal S}'_2,a',b')$, (resp. {\small H}$({\cal S}'_1,a',b)$ and {\small H}$({\cal S}'_2,a',b')$).
We need to check that $a$ and $a'$ (resp. $b$ and $b'$) do not become dangerous in ${\cal C}_{n+2}$ (if they were not in ${\cal C}_n$) and 
that the new edges and new vertex do not contradict the claim. 
There are now, unfortunately, 8 cases we need to inspect, depending on $f$ and $f'$:

\noindent\textbullet
{\bf $f,f'$ of the form \ding{195}}. Then each of ${\cal S}_1$, ${\cal S}_2$, ${\cal S}'_1$, ${\cal S}'_2$ is of the form $\spg{i}{}$ for some $i\neq \alpha,\eta_0, \eta_1$. By assumption none of $a, a',b$ is dangerous in ${\cal C}_n$ and they remain non-dangerous in ${\cal C}_{n+2}$. The new $b'$ is non-dangerous either. The claim holds in ${\cal C}_{n+2}$.  

\noindent\textbullet
{\bf $f,f'$ of the form \ding{196}}. Analogous to the previous case.

\noindent\textbullet
{\bf $f,f'$ of the form \ding{192}A, \ding{192}B}. Then ${\cal S}_1={\cal S}_2=\spg{}{}$ so  $a,a'$ and $b$ are all dangerous. 
${\cal S}'_1=\spg{\alpha}{}$ and ${\cal S}'_2=\spg{\eta_1}{}$ and the new $b'$ is non-dangerous in ${\cal C}_{n+2}$. The claim holds in ${\cal C}_{n+2}$. 

\noindent\textbullet
{\bf $f,f'$ of the form \ding{192}B, \ding{192}A}. Then ${\cal S}_1=\spg{\alpha}{}$ and ${\cal S}_2=\spg{\eta_1}{}$ so, by assumption, $a$ and $a'$ must have already been dangerous in ${\cal C}_n$.  ${\cal S}'_1={\cal S}'_2=\spg{}{}$ and so the new $b'$ is created as dangerous. The claim holds in ${\cal C}_{n+2}$.

\noindent\textbullet
{\bf $f,f'$ of the form \ding{193}A, \ding{193}B}. Then ${\cal S}_1=\spg{\eta_0}{}$ and ${\cal S}_2=\spg{}{}$. By assumption $b$ and $a'$ are dangerous but
$a$ is not. ${\cal S}'_1=\spg{\beta_0}{}$ and ${\cal S}_2=\spg{\eta_1}{}$, so $b'$ is non-dangerous and $a$ remains non-dangerous in ${\cal C}_{n+2}$.
The claim holds in ${\cal C}_{n+2}$.

\noindent\textbullet
{\bf $f,f'$ of the form \ding{193}B, \ding{193}A}. ${\cal S}'_1=\spg{\beta_0}{}$ and ${\cal S}_2=\spg{\eta_1}{}$. 
By assumption $a'$ is dangerous while $a$ and $b$ are not.  ${\cal S}'_1=\spg{\eta_0}{}$ and ${\cal S}_2=\spg{}{}$ so $b'$ is created as dangerous
but $a$ remains non-dangerous in ${\cal C}_{n+2}$.
The claim holds in ${\cal C}_{n+2}$.

\noindent
The {\bf two cases with \ding{194}} are analogous to the cases with \ding{193}.\eop

\section{Characterization of $W({\cal C}^{{\cal Q}_0})$}\label{bezczerwone} 

A word $i_0i_1\ldots i_{l-1}i_l\in {\mathbb S}^*$ is {\em correct} if 
for all $k\neq 0$ there is $i_k\neq \alpha$ and for all  $k\neq l$ there is $i_k\neq \eta_0$ and $i_k\neq \eta_1$. 
A word $i_0i_1\ldots i_{l-1}i_l\in {\mathbb S}^*$ is {\em maximal correct} if it is correct and $i_0=\alpha$ and  $i_l=\eta_0$ or $i_l= \eta_1$

\lemmaaa{\label{tylkothue}
(i) For each correct  ${\bf w}\in W({\cal C}^{{\cal Q}_0})$ there is a maximal 
correct  ${\bf v}\in W({\cal C}^{{\cal Q}_0})$  such that $\bf w$ is a subword of $\bf v$. 

(ii) If  ${\bf v}\in W({\cal C}^{{\cal Q}_0})$ is maximal correct then ${\bf v} \stackrel{\ast\;\;\;}{\Leftrightarrow_{\Pi}} \alpha\eta_1$.
}

\noindent {\em Proof:} It is enough to prove that both claims hold in each $W({\cal C}_n)$, and this can be proved by induction.
 The claim is clearly true in ${\cal C}_0$ as $W({\cal C}_0)$ is empty.
The induction step follows the proof of Lemma \ref{niebezpieczne}, and similar case inspection is needed. 
 For {\bf $f,f'$ of the form \ding{195}} apply the argument from the proof of Lemma \ref{thuedziala}. 

Also for 
both the \ding{193} and both the \ding{194} cases the validity of the induction hypothesis for $n+2$ 
follows from the assumption about $W({\cal C}_{n})$ and the fact that the  word ${\bf w}\in W({\cal C}_{n+2})$ 
 under consideration is a result of one rewriting, using one of the rules from $\Pi_<$, and applied to some word in ${\cal C}_{n}$.

In the case of {\bf $f,f'$ of the form \ding{192}B, \ding{192}A} no new words are added to $W({\cal C}_{n+2})$. 
Finally, in the case of {\bf $f,f'$ of the form \ding{192}A, \ding{192}B} one new correct 
word is created\footnote{What is actually created is a new copy of this word.}. It is 
$\alpha\eta_1$, which clearly satisfies both the claims of the Lemma.\eop

Now it easily follows from Lemma \ref{tylkothue} that:

\lemmaaa{\label{bezgammy}
No edge in ${\cal C}^{{\cal Q}_0}$ is labeled with $\spg{\gamma}{}$ or with $\spg{\gamma'}{}$
}

\section{From ${\cal C}^{{\cal Q}_0}$ to ${\cal C}^{{\cal Q}}$}\label{ostatnia}

For each pair of edges\footnote{
In formal terms this means that we extend the fair sequence  $\{{\cal C}_n\}_{n\in \omega}$ with new 
structures. We do not rely on that so we do not need to prove it, but there are infinitely many of the new structures,
 as there are infinitely many edges in ${\cal C}^{{\cal Q}_0}$ labeled with $\spg{}{}$. Thus the new fair sequence is
$\{{\cal C}_n\}_{n\in 2\omega}$
 The structure $\cal C$ is now -- as always -- defined as $\bigcup_{n\in 2\omega} {\cal C}_n$.
}
 of the form {\small H}$(\spg{}{},a,b)$, {\small H}$(\spg{}{},a,b')$  in ${\cal C}^{{\cal Q}_0}$ 
let us now apply a rewriting using the rule \ding{197}. Each such rewriting will result in adding a new vertex $a'$ and 
new edges {\small H}$(\spr{\gamma}{},a',b)$ and {\small H}$(\spr{\gamma'}{r},a',b')$. Call the resulting structure $\cal C$.
Notice that all the new vertices of $\cal C$ are of degree 2.

{\bf Proof of} Lemma \ref{usmieszki}, and thus of Lemma \ref{main2} and {\bf Theorem \ref{main}, will be completed once we show}:

\lemmaaa{\label{drugieomega}
(i) ToDo$(Q,{\cal C})$ is empty. In consequence $\cal C$=${\cal C}^{{\cal Q}}$. 
(ii) There is no edge labeled with $\spr{}{}$ in $\cal C$.
}

\noindent {\em Proof:} Claim (ii) is obvious -- there was no such edge in ${\cal C}^{{\cal Q}_0}$ and we never added one while building $\cal C$ on the top of 
${\cal C}^{{\cal Q}}$. 
For the proof of Claim (i) first notice that no rewriting with green inputs is possible in $\cal C$: 

--no such rewriting using rules from
${\cal Q}_0$ is possible since $\cal C$ has no new green edges compared to ${\cal C}^{{\cal Q}_0}$ and 

-- no such rewriting 
using rule \ding{197} and at least one 1-lame green edge is possible, since no edge of ${\cal C}^{{\cal Q}_0}$ is labeled with $\spg{\gamma}{}$ or
$\spg{\gamma'}{}$ (Lemma \ref{bezgammy}), and

-- no such rewriting using rule \ding{197} and both inputs labeled with $\spg{}{}$ is possible any more -- by the definition of  $\cal C$.

Now how about the possibility of rewritings in $\cal C$ using red edges as the input? No such rewriting using rules of ${\cal Q}_0$ and 
having, as the input, at least one red edge from  ${\cal C}^{{\cal Q}_0}$ is possible, by Lemma \ref{wmalzenstwie} (however tempting the new red edges would look!). 
By Lemma \ref{czerwonedolne} 
neither 
$\fff^\gamma$ nor $\fff^{\gamma'}_{r}$ match with any red edge from from  ${\cal C}^{{\cal Q}_0}$. This means that no red edge 
from   ${\cal C}^{{\cal Q}_0}$ can be an input of any new rewriting in $\cal C$. 

To finish the proof notice that none of the  rewritings from ${\mathbb F}^\oaa_s$ can use either $\spr{\gamma}{}$ or
$\spr{\gamma'}{r}$ as one of its inputs. Since all the new edges in $\cal C$ are of degree 2 the only rule from ${\mathbb F}^\obb_s$ that 
matches with the new edges of $\cal C$ is \ding{197}, which however cannot be used due to idempotence reasons.  
\eop

\bibliographystyle{alpha}
\bibliography{determinacja6}

\begin{thebibliography}{FKN13}

\bibitem[AD98]{AD98}
Serge Abiteboul and Oliver~M. Duschka.
\newblock Complexity of answering queries using materialized views.
\newblock In {\em Proceedings of the Seventeenth ACM SIGACT-SIGMOD-SIGART
  Symposium on Principles of Database Systems}, PODS '98, pages 254--263, New
  York, NY, USA, 1998. ACM.

\bibitem[Afr11]{A11}
Foto~N. Afrati.
\newblock Determinacy and query rewriting for conjunctive queries and views.
\newblock {\em Theor. Comput. Sci.}, 412(11):1005--1021, March 2011.

\bibitem[Dav77]{D77}
M.~Davis.
\newblock Unsolvable problems.
\newblock In J.~Barwise, editor, {\em Handbook of Mathematical Logic}, pages
  567--594. North-Holland, Amsterdam, 1977.

\bibitem[DPT99]{DPT99}
Alin Deutsch, Lucian Popa, and Val Tannen.
\newblock Physical data independence, constraints, and optimization with
  universal plans.
\newblock In Malcolm~P. Atkinson, Maria~E. Orlowska, Patrick Valduriez,
  Stanley~B. Zdonik, and Michael~L. Brodie, editors, {\em VLDB'99, Proceedings
  of 25th International Conference on Very Large Data Bases, September 7-10,
  1999, Edinburgh, Scotland, UK}, pages 459--470. Morgan Kaufmann, 1999.

\bibitem[FG12]{FG12}
Enrico Franconi and Paolo Guagliardo.
\newblock The view update problem revisited.
\newblock {\em CoRR}, abs/1211.3016, 2012.

\bibitem[FGZ12]{FGZ12}
Wenfei Fan, Floris Geerts, and Lixiao Zheng.
\newblock View determinacy for preserving selected information in data
  transformations.
\newblock {\em Inf. Syst.}, 37(1):1--12, March 2012.

\bibitem[FKN13]{FKN13}
Enrico Franconi, Volha Kerhet, and Nhung Ngo.
\newblock Exact query reformulation over databases with first-order and
  description logics ontologies.
\newblock {\em J. Artif. Intell. Res. {(JAIR)}}, 48:885--922, 2013.

\bibitem[Hal01]{H01}
Alon~Y. Halevy.
\newblock Answering queries using views: A survey.
\newblock {\em The VLDB Journal}, 10(4):270--294, December 2001.

\bibitem[JK82]{JK82}
D.~S. Johnson and A.~Klug.
\newblock Testing containment of conjunctive queries under functional and
  inclusion dependencies.
\newblock In {\em Proceedings of the 1st ACM SIGACT-SIGMOD Symposium on
  Principles of Database Systems}, PODS '82, pages 164--169, New York, NY, USA,
  1982. ACM.

\bibitem[LMS95]{LMS95}
Alon~Y. Levy, Alberto~O. Mendelzon, and Yehoshua Sagiv.
\newblock Answering queries using views (extended abstract).
\newblock In {\em Proceedings of the Fourteenth ACM SIGACT-SIGMOD-SIGART
  Symposium on Principles of Database Systems}, PODS '95, pages 95--104, New
  York, NY, USA, 1995. ACM.

\bibitem[LY85]{LY85}
Per-Ake Larson and H.~Z. Yang.
\newblock Computing queries from derived relations.
\newblock In {\em Proceedings of the 11th International Conference on Very
  Large Data Bases - Volume 11}, VLDB '85, pages 259--269. VLDB Endowment,
  1985.

\bibitem[NSV07]{NSV07}
Alan Nash, Luc Segoufin, and Victor Vianu.
\newblock Determinacy and rewriting of conjunctive queries using views: A
  progress report.
\newblock In Thomas Schwentick and Dan Suciu, editors, {\em Database Theory –
  ICDT 2007}, volume 4353 of {\em Lecture Notes in Computer Science}, pages
  59--73. Springer Berlin Heidelberg, 2007.

\bibitem[NSV10]{NSV10}
Alan Nash, Luc Segoufin, and Victor Vianu.
\newblock Views and queries: Determinacy and rewriting.
\newblock {\em ACM Trans. Database Syst.}, 35:21:1{\textendash}21:41, July
  2010.

\bibitem[Pas11]{P11}
Daniel Pasail{\u{a}}.
\newblock Conjunctive queries determinacy and rewriting.
\newblock In Tova Milo, editor, {\em {P}roceedings of the 14th {I}nternational
  {C}onference on {D}atabase {T}heory ({ICDT}'11)}, pages 220--231, Uppsala,
  Sweden, March 2011. ACM Press.

\bibitem[SV05]{SV05}
Luc Segoufin and Victor Vianu.
\newblock Views and queries: Determinacy and rewriting.
\newblock In {\em Proceedings of the Twenty-fourth ACM SIGMOD-SIGACT-SIGART
  Symposium on Principles of Database Systems}, PODS '05, pages 49--60, New
  York, NY, USA, 2005. ACM.

\bibitem[YL87]{YL87}
H.~Z. Yang and Per-Ake Larson.
\newblock Query transformation for psj-queries.
\newblock In {\em Proceedings of the 13th International Conference on Very
  Large Data Bases}, VLDB '87, pages 245--254, San Francisco, CA, USA, 1987.
  Morgan Kaufmann Publishers Inc.

\end{thebibliography}

%%%%%%%%%%%%%%%%%%%%%%%%%%%%%%%%%%%%%%%%%%%%%%%%%%%%%%%%%%%%%%%%%%%%%%%%%%%%%%%%%%%%%%%%%%%%%%%%%%%%%%%
%%%%%%%%%%%%%%%%%%%%%%%%%%%%%%%%%%%%%%%%%%%%%%%%%%%%%%%%%%%%%%%%%%%%%%%%%%%%%%%%%%%%%%%%%%%%%%%%%%%%%%%
\end{document}